%% file: main.tex
\documentclass[letterpaper,twocolumn,10pt]{article}
\usepackage{usenix-16-2-23}

\usepackage{preamble}

\input{header}

\input{graphs}

\pdfoutput=1

\newcommand{\nc}[1]{{\color{cyan}\grumbler{Natacha}{#1}}}
\newcommand{\sg}[1]{{\color{brown}\grumbler{Suyash}{#1}}}

\newcommand{\raf}[1]{{\color{green}\grumbler{Reggie}{#1}}}
\newcommand{\mk}[1]{{\color{blue}\grumbler{Manos}{#1}}}

\title{Picsou: Enabling Replicated State Machines to Communicate Efficiently}

\author{
Reginald Frank,
Micah Murray, \\
Chawinphat Tankuranand,
Junseo Yoo,
Ethan Xu,\\
Natacha Crooks,
Suyash Gupta\textsuperscript{$\dagger$}, 
Manos Kapritsos\textsuperscript{*}\\
\small University of California, Berkeley;
\small \textsuperscript{$\dagger$}University of Oregon;
\small \textsuperscript{*}University of Michigan\\
\vspace{2cm}
}

\begin{document}

\maketitle

\begin{abstract}
 Replicated state machines (\RSM{s}) cannot communicate effectively today as there is no formal framework or efficient protocol to do so. 
To address this issue, we introduce a new primitive, \CCCFull{} (\CCC{}) and present \Scrooge{}, a practical implementation of the \CCC{} primitive.
\Scrooge{} draws inspiration from networking and TCP to allow two \RSM{s} to communicate with constant metadata overhead in the failure-free case and a minimal number of message resends in the case of failures. 
\Scrooge{} is flexible and allows both crash fault tolerant and Byzantine fault tolerant consensus protocols to communicate. At the heart of \Scrooge{}'s good performance and generality is the concept of \quack{}s (quorum acknowledgments). \quack{}s allow nodes in each \RSM{} to precisely determine when messages have definitely been received, or likely lost. Our results are promising: we obtain up to $24\times$ better performance than prior solutions \revision{on microbenchmarks and applications, ranging from disaster recovery to data reconciliation}{}. 
\end{abstract}

\input{intro-new}

\input{prelim}

\input{primitive}
\input{algo-nc}

\input{eval-micah}

\input{related}

\input{conclusion}

\bibliographystyle{plain}
\bibliography{refined}

\appendix
\input{proofs}

\end{document}

%% file: header.tex
\DeclareSIUnit{\txn}{txn}
\DeclareSIUnit{\batch}{batch}
\newcommand{\Name}[1]{\textsc{#1}}
\newcommand{\BFT}{\textsc{BFT}}
\newcommand{\CFT}{\textsc{CFT}}

\newcommand{\RSM}{\textsc{RSM}}
\newcommand{\PoS}{\textsc{PoS}}
\newcommand{\pbft}{\textsc{PBFT}}

\newcommand{\Scrooge}{\textsc{Picsou}}
\newcommand{\quack}{\textsc{Quack}}

\newcommand{\Replica}[2]{\textsc{R}_{{#1}{#2}}}
\newcommand{\SMR}[1]{\mathcal{R}_{#1}}

\newcommand{\CCC}{\textsc{C3B}}
\newcommand{\Transaction}[1][t]{\MakeUppercase{#1}} 

\newcommand{\CCCFull}{Cross-Cluster Consistent Broadcast}

\newcommand{\n}[1]{\textbf{n}_{#1}}
\newcommand{\f}[1]{\textbf{f}_{#1}}
\newcommand{\uf}[1]{\textbf{u}_{#1}}
\newcommand{\rf}[1]{\textbf{r}_{#1}}
\newcommand{\alp}[1]{\textbf{a}_{#1}}

\newcommand{\share}[1]{{\delta}_{#1}}
\newcommand{\Ack}[1]{\Name{Ack}(#1)}
\newcommand{\AckPlain}{\Name{Ack}}

\newcommand{\SignMessage}[2]{\langle#1\rangle_{#2}}
\newcommand{\Seqn}{k}
\newcommand{\Qusign}[1]{\mathcal{Q}_{#1}}

\newcommand{\ETCD}{\textsc{ETCD}}
\DeclareMathOperator{\lcm}{lcm}

\newcommand{\abs}[1]{\lvert #1 \rvert}

\newcommand{\LL}{\textsc{LL}}
\newcommand{\ATA}{\textsc{ATA}}
\newcommand{\OTO}{\textsc{OST}}
\newcommand{\OTF}{\textsc{OTU}}

\newcommand{\Kafka}{\textsc{Kafka}}

\newcommand{\ResDB}{ResilientDB}
\newcommand{\Algo}{Algorand}
\newcommand{\File}{File}
\newcommand{\Raft}{Raft}


%% file: graphs.tex
\usepackage{xcolor}
\usepackage{tikz,pgfplots,pgfplotstable}
\usetikzlibrary{arrows.meta}
\tikzset{
    >=Stealth,
    plot/.append style={baseline,scale=0.45},
    label/.append style={font=\strut\footnotesize},
    dot/.style={circle,scale=0.35,draw=black,fill=black},
}
\pgfplotscreateplotcyclelist{mycyclelist}{
    red   ,every mark/.append style={solid,fill=\pgfplotsmarklistfill},mark=*\\
    blue!90!white!90     ,every mark/.append style={solid,fill=\pgfplotsmarklistfill},mark=square*\\
    blue    ,every mark/.append style={solid,fill=\pgfplotsmarklistfill},mark=triangle*\\
    black    ,every mark/.append style={solid,fill=\pgfplotsmarklistfill},mark=pentagon*\\
    brown   ,every mark/.append style={solid,fill=\pgfplotsmarklistfill},mark=o\\
    orange  ,every mark/.append style={solid,fill=\pgfplotsmarklistfill},mark=halfcircle*\\
    teal  ,every mark/.append style={solid,fill=\pgfplotsmarklistfill,rotate=180},mark=halfdiamond*\\
    green  ,every mark/.append style={solid,fill=\pgfplotsmarklistfill,rotate=180},mark=diamond*\\
}
\pgfplotscreateplotcyclelist{mycyclelistex}{
    black   ,every mark/.append style={solid,fill=\pgfplotsmarklistfill},mark=*\\
    red     ,every mark/.append style={solid,fill=\pgfplotsmarklistfill},mark=square*\\
}
\pgfplotsset{
    tick label style={font=\Large},
    legend style={font=\large,cells={anchor=west},text=black},
    title style={font=\Large},
    label style={font=\normalsize},
    width=215pt,
    height=185pt,
    every axis/.append style={
            ylabel near ticks,
            xlabel near ticks,
            mark size=3pt,
            cycle list name=mycyclelist,
            font=\Large,
            y tick label style={
                    /pgf/number format/precision=0,
                    /pgf/number format/fixed,
                    /pgf/number format/fixed zerofill
                }
        },
}

\pgfplotstableread{
 X      Picsou             ATA                 OTO                 GeoBFT              LL                  KFK
 0.1    7738644.743780447   3081658.2872693     12866092.98794477   3965800.19237398    4686759.090565393   694938.3374999999
 1      1142557.996410039   431415.93674443     3480954.879824384   437100.786441546    513055.4397405473   50359.521875
 10     83252.40331838577   42388.607971712     390472.0903338299   43933.1444264892    50862.17991966786   10712.703124999998
 100    9025.931034636726   4128.5553517183     42980.16148973186   4351.05145035048    5237.773824722042   672.971875
 1000   1426.894813082035   440.67502195625     3540.400121813411   435.850399947455    521.5165186712587   49.090625
}\graphFFtp

\pgfplotstableread{
 X      Picsou             ATA                 OTO                 GeoBFT              LL                  KFK
 0.1    4345095.414509543   659710.5590991525   21936451.39290192   956202.5350474387   987378.2626495629   147372.1302631579
 1      754481.4527517064   105382.2792080013   18056792.93564376   101236.3048067124   107403.9947530283   26683.739473684214
 10     72722.56881176609   10744.39903298061   1888899.156479316   10163.78391941899   10841.60397396319   1341.238157894737
 100    8551.882223489463   1080.444843424772   188081.6070283802   1052.350146348354   1080.53326132012    211.1315789473684
 1000   1302.7996743003077   107.2600150169021   18298.13926261384   108.0000234000053   108.5498895927236   119.95131578947363
}\graphFStp

\pgfplotstableread{
 X  Picsou             ATA                 OTO                 GeoBFT              LL                  KFK
 4  7738644.743780447   3081658.2872693     12866092.98794477   3965800.19237398	   4686759.090565393   425379.04375
 7  7565866.825638401   1878357.9391401     16508537.55108685   2538811.54567052    2731381.854340263   321501.6875
 10 7564097.396889148   1231317.3160471     18513127.04284093   1779033.59152683    1894406.786488369   250706.64500000002
 13 6864386.840522587   973464.68595800     19951769.39428246   1337819.27521096    1395529.284675436   209206.2432692307
 16 4168019.063695851   794715.68732564     21173899.44184122   1048861.00776384    1152269.656675954   173840.74843749998
 19 4345095.414509543   659710.55909915     21936451.39290192   906622.430187471    995486.0827307655   141144.30065789475
}\graphFmsgB

\pgfplotstableread{
 X  Picsou             ATA                 OTO                 GeoBFT              LL                  KFK
 4  1426.894813082035   440.6750219562555   3540.400121813411   422.099992905001    514.099735371958    47.059374999999996
 7  1318.5503518951073   276.1409207172276   6628.510198855307   282.433219044728    298.166199262672    27.12142857142857
 10 1481.1003209050732   198.0900318220717   9107.801063935844   211.416687404446    206.499882983438    18.16
 13 1356.6339664291843   153.9300254709225   12538.2445666999   155.449558516816    158.5499762211221   13.030769230769229
 16 1417.100070855004   126.5149904580530   15111.0433251365   123.049997776382    129.166698958349    7.640624999999999
 19 1302.7996743003077   107.2600150169021   18298.1392626138   104.050054088368    108.916267351089    7.730147058823529
}\graphFmsgMB

\pgfplotstableread{
 X  PicsouGrad       PicsouUnfair    PicsouFair        PicsouLarge      
 4  1223.422526765891 993.612834783762 105.69999656000023 1649.520309562079 
 7  1239.025015181898 591.287596122520 104.69998509334587 1667.166528421678 
 10 1126.862812751358 383.112533510629 76.16668948779224  1699.684205286041 
 13 1099.860038509602 305.999850744200 89.3000304216826   1664.366754073342 
 16 1031.515084390959 248.975028118441 48.43330666669511  1499.733826560615 
 19 1087.779003701413 194.099968458757 44.01666936891273  1431.516852120708 
}\graphStakeOld

\pgfplotstableread{
 X  Picsou1   Picsou2   Picsou4   Picsou8 Picsou16 Picsou32 Picsou64
 4  6816474.93 6734572.06 6410692.55 5848139.31 5413773.62 4919940.22 4699746.85 
 7  6885099.77 6719525.53 6117926.84 4710647.10 3531858.08 3132212.34 2743074.05
 10 7102860.45 6739133.58 5794461.06 3951459.59 2930689.74 2417113.17 2122422.72 
 13 7023094.09 6980671.92 5044135.44 3594480.84 2500937.93 1982428.60 1705959.26 
 16 5779183.23 5721700.63 4017465.56 3247222.25 2249168.53 1718583.90 1427907.43
 19 4814106.24 5054078.37 3311041.91 2913396.14  2000543.67 1455260.31 1212767.64 
}\graphStakeNewNoThrottle

\pgfplotstableread{
 X  Picsou1   Picsou2   Picsou4   Picsou8 Picsou16 Picsou32 Picsou64
 4 999475.08 999934.42 999948.78 999911.88 999794.40 999891.05 999963.00  
 7   999652.55 999652.55 999490.75 999647.23 999869.37 999852.48 999707.55
 10  999069.87 999360.72 999525.70 999629.73 999596.47 999723.68 999493.33
 13  999615.00 999496.65 999353.17 999457.37 999531.92 999464.15 995340.15
 16  998942.42 998807.52 998599.18 998892.55 998849.60 998803.98 990156.66 
 19 999006.23 998812.73 999086.08 965856.93 998957.20 969259.23 934584.37
}\graphStakeNewThrottle

\pgfplotstableread{
 X       Picsou    ATA         OTO     GeoBFT  LL      KFK ETCD
 .245    19.96      14.51       21.05   17.67   19.53   20.17   18.80
 .498    36.04      30.74       37.16   35.56   35.45   38.10   35
 1.980   65.39      46.88       65.38   45.94   46.47   31.79   68.17
 4.020   66.39      49.39       66.41   45.74   46.83   39.86   69.28
 14.304  63.14      45.75       63.77   44.98   47.58   54.40   68.26
}\graphDRPlot

\pgfplotstableread{
 X  Picsou         ATA     OTO     GeoBFT      LL      KFK ETCD
 .245   1.24        1.00    13.82   3.18        2.11    1.5304837663968405   18.88
 .498   2.01        2.76    25.15   1.97        3.60    2.5818304777145387   35
 1.980  26.52       10.81   33.61   17.78       7.69    11.806285772323609   68.17
 4.020  46.58       27.64   61.71   37.12       34.15   17.211844720840453   69.28
 8.052  62.95       49.26   66.88   49.49       49.59   26.920806058247884   67.21
 14.304 62.90       49.32   63.64   49.79       49.65   46.94892041524251   68.26
}\graphCCFPlot

\pgfplotstableread{
 X  ATA  Picsou OTO OTU LL KFK
 4  440.6750219562555 1100.6841228821827 2992.25 382.68 490.06 54.395
 7  276.1409207172276   1012.3503364790188 4269.92 233.37 294.43 33.3222
 10 198.0900318220717   1029.7852803799906 6575.80 159.43 196.93 19.078
 13 153.9300254709225   1035.4707149424196 9373.85 122.23  148.60 12.161
 16 126.5149904580530   1020.0827540731508 10118.36 98.37 113.57 10.126
 19 107.2600150169021   959.7336372195408 11705.97 64.60 71.03 8.961
}\graphMaxFail

\pgfplotstableread{
 X  ATA                 OTO                 Picsou       OTU       LL
 4  22.90000047600006   89.599              272.084969   22.349944  22.000006
 10 23.175006           223.799996          928.12004   22.283306    22.100039
 19 23.440001        424.269985   1018.99006            22.550014     22.233366
}\graphGeo

\pgfplotstableread{
 X  ATA                 Picsou-99          Picsou-0           Picsou-ack-delay
 4  440.6750219562555   1470.5178240286596  1484.0329870589696  1274.8667959466925
 7  276.1409207172276   1194.0340728646145  1324.9506919386286  1520.3333469333757
 10 198.0900318220717   1328.8002723790069  1433.2659500336915  1361.216014963024
 13 153.9300254709225   1433.1673354781472  1274.1003397600907  1631.282605274769
 16 126.5149904580530   1574.4398610266198  1520.8007604003803  1562.7002344050356
 19 107.2600150169021   1522.9655370073674  1542.133256182237   1367.608673303876
}\graphByzAck

\pgfplotstableread{
 X  P0                  P64                   P128                P192                P256              
 4  975.0838837221744   1012.7336017290004    1077.016953934564   1100.68412288217    1080.5847006250328
 7  894.5166369413357   877.5333918355605     999.33329927113     1002.783816844129   1012.3503364790188
 10 493.08343195001976  483.8167875100368     955.200222880053    990.9501486367026   1029.7852803799906
 13 352.1833214494511   354.2668142778402     734.8163962425576   992.5669975223336   1035.4707149424196
 16 277.76668518444603  274.43337907222985    565.4667514255722   885.0845700932289   1020.0827540731508
 19 232.6005177889451   234.2501556584953     436.5833478777787   697.4503255018379   959.7336372195408
}\graphByzPhiList

\newcommand{\axisGeoTP}{Throughput (\si{\txn\per\second})}

\newcommand{\axisMsg}{Message size (\SI{}{kB})}
\newcommand{\axisReps}{Replicas (per \RSM{})}

\newcommand{\graphFMsgB}{
    \begin{tikzpicture}[plot]
        \begin{axis}[
                title={(i) Message Size = \SI{0.1}{kB}},
                scaled y ticks = false,
                xlabel={\axisReps},
                ylabel={\axisGeoTP},
                ymin=0,
                xtick={4,7,10,13,16,19,22},
                ymode=log,
                log ticks with fixed point,
                legend to name={mainlegend},
                legend columns=-1,
            ]

            \addplot [black,every mark/.append style={solid,fill=\pgfplotsmarklistfill},mark=pentagon*] table[x={X},y={Picsou}] {\graphFmsgB};
            \addplot [blue!90!white!90,every mark/.append style={solid,fill=\pgfplotsmarklistfill},mark=o] table[x={X},y={ATA}] {\graphFmsgB};
            \addplot [red,every mark/.append style={solid,fill=\pgfplotsmarklistfill},mark=halfdiamond*] table[x={X},y={OTO}] {\graphFmsgB};
            \addplot [green!50!black!70,every mark/.append style={solid,fill=\pgfplotsmarklistfill},mark=oplus] table[x={X},y={GeoBFT}] {\graphFmsgB};
            \addplot [orange,every mark/.append style={solid,fill=\pgfplotsmarklistfill},mark=triangle] table[x={X},y={LL}] {\graphFmsgB};
            \addplot [orange!30!black!80,every mark/.append style={solid,fill=\pgfplotsmarklistfill},mark=square] table[x={X},y={KFK}] {\graphFmsgB};

            \addlegendentry [black,every mark/.append style={solid,fill=\pgfplotsmarklistfill},mark=pentagon*] \Scrooge
            \addlegendentry [blue!90!white!90,every mark/.append style={solid,fill=\pgfplotsmarklistfill},mark=o] \ATA
            \addlegendentry [red,every mark/.append style={solid,fill=\pgfplotsmarklistfill},mark=halfdiamond*] \OTO
            \addlegendentry [green!50!black!70,every mark/.append style={solid,fill=\pgfplotsmarklistfill},mark=oplus] \OTF
            \addlegendentry [orange,every mark/.append style={solid,fill=\pgfplotsmarklistfill},mark=triangle] \LL
            \addlegendentry [orange!30!black!80,every mark/.append style={solid,fill=\pgfplotsmarklistfill},mark=square] \Kafka
        \end{axis}
    \end{tikzpicture}
}

\newcommand{\graphFMsgMB}{
    \begin{tikzpicture}[plot]
        \begin{axis}[
                title={(ii) Message Size = \SI{1}{MB}},
                scaled y ticks = false,
                xlabel={\axisReps},
                ylabel={\axisGeoTP},
                ymin=0,
                xtick={4,7,10,13,16,19,22},
                ymode=log,
                log ticks with fixed point,
                legend to name={mainlegend},
                legend columns=-1,
            ]

            \addplot [black,every mark/.append style={solid,fill=\pgfplotsmarklistfill},mark=pentagon*] table[x={X},y={Picsou}] {\graphFmsgMB};
            \addplot [blue!90!white!90,every mark/.append style={solid,fill=\pgfplotsmarklistfill},mark=o] table[x={X},y={ATA}] {\graphFmsgMB};
            \addplot [red,every mark/.append style={solid,fill=\pgfplotsmarklistfill},mark=halfdiamond*] table[x={X},y={OTO}] {\graphFmsgMB};
            \addplot [green!50!black!70,every mark/.append style={solid,fill=\pgfplotsmarklistfill},mark=oplus] table[x={X},y={GeoBFT}] {\graphFmsgMB};
            \addplot [orange,every mark/.append style={solid,fill=\pgfplotsmarklistfill},mark=triangle] table[x={X},y={LL}] {\graphFmsgMB};
            \addplot [orange!30!black!80,every mark/.append style={solid,fill=\pgfplotsmarklistfill},mark=square] table[x={X},y={KFK}] {\graphFmsgMB};

            \addlegendentry [black,every mark/.append style={solid,fill=\pgfplotsmarklistfill},mark=pentagon*] \Scrooge
            \addlegendentry [blue!90!white!90,every mark/.append style={solid,fill=\pgfplotsmarklistfill},mark=o] \ATA
            \addlegendentry [red,every mark/.append style={solid,fill=\pgfplotsmarklistfill},mark=halfdiamond*] \OTO
            \addlegendentry [green!50!black!70,every mark/.append style={solid,fill=\pgfplotsmarklistfill},mark=oplus] \OTF
            \addlegendentry [orange,every mark/.append style={solid,fill=\pgfplotsmarklistfill},mark=triangle] \LL
            \addlegendentry [orange!30!black!80,every mark/.append style={solid,fill=\pgfplotsmarklistfill},mark=square] \Kafka
        \end{axis}
    \end{tikzpicture}
}

\newcommand{\graphFFTP}{
    \begin{tikzpicture}[plot]
        \begin{axis}[
                title={(iii) $\n{} = 4$ replicas},
                scaled y ticks = false,
                xlabel={\axisMsg},
                ylabel={\axisGeoTP},
                ymin=0,
                ymode=log,
                xmode=log,
                log ticks with fixed point,
                legend to name={mainlegend},
                legend columns=-1,
            ]

            \addplot [black,every mark/.append style={solid,fill=\pgfplotsmarklistfill},mark=pentagon*] table[x={X},y={Picsou}] {\graphFFtp};
            \addplot [blue!90!white!90,every mark/.append style={solid,fill=\pgfplotsmarklistfill},mark=o] table[x={X},y={ATA}] {\graphFFtp};
            \addplot [red,every mark/.append style={solid,fill=\pgfplotsmarklistfill},mark=halfdiamond*] table[x={X},y={OTO}] {\graphFFtp};
            \addplot [green!50!black!70,every mark/.append style={solid,fill=\pgfplotsmarklistfill},mark=oplus] table[x={X},y={GeoBFT}] {\graphFFtp};
            \addplot [orange,every mark/.append style={solid,fill=\pgfplotsmarklistfill},mark=triangle] table[x={X},y={LL}] {\graphFFtp};
            \addplot [orange!30!black!80,every mark/.append style={solid,fill=\pgfplotsmarklistfill},mark=square] table[x={X},y={KFK}] {\graphFFtp};

            \addlegendentry [black,every mark/.append style={solid,fill=\pgfplotsmarklistfill},mark=pentagon*] \Scrooge
            \addlegendentry [blue!90!white!90,every mark/.append style={solid,fill=\pgfplotsmarklistfill},mark=o] \ATA
            \addlegendentry [red,every mark/.append style={solid,fill=\pgfplotsmarklistfill},mark=halfdiamond*] \OTO
            \addlegendentry [green!50!black!70,every mark/.append style={solid,fill=\pgfplotsmarklistfill},mark=oplus] \OTF
            \addlegendentry [orange,every mark/.append style={solid,fill=\pgfplotsmarklistfill},mark=triangle] \LL
            \addlegendentry [orange!30!black!80,every mark/.append style={solid,fill=\pgfplotsmarklistfill},mark=square] \Kafka
        \end{axis}
    \end{tikzpicture}
}

\newcommand{\graphFSTP}{
    \begin{tikzpicture}[plot]
        \begin{axis}[
                title={(iv) $\n{} = 19$ replicas},
                scaled y ticks = false,
                xlabel={\axisMsg},
                ylabel={\axisGeoTP},
                ymin=0,
                ymode=log,
                xmode=log,
                log ticks with fixed point,
                legend to name={mainlegendN},
                legend columns=-1,
            ]

            \addplot [black,every mark/.append style={solid,fill=\pgfplotsmarklistfill},mark=pentagon*] table[x={X},y={Picsou}] {\graphFStp};
            \addplot [blue!90!white!90,every mark/.append style={solid,fill=\pgfplotsmarklistfill},mark=o] table[x={X},y={ATA}] {\graphFStp};
            \addplot [red,every mark/.append style={solid,fill=\pgfplotsmarklistfill},mark=halfdiamond*] table[x={X},y={OTO}] {\graphFStp};
            \addplot [green!50!black!70,every mark/.append style={solid,fill=\pgfplotsmarklistfill},mark=oplus] table[x={X},y={GeoBFT}] {\graphFStp};
            \addplot [orange,every mark/.append style={solid,fill=\pgfplotsmarklistfill},mark=triangle] table[x={X},y={LL}] {\graphFStp};
            \addplot [orange!30!black!80,every mark/.append style={solid,fill=\pgfplotsmarklistfill},mark=square] table[x={X},y={KFK}] {\graphFStp};

            \addlegendentry [black,every mark/.append style={solid,fill=\pgfplotsmarklistfill},mark=pentagon*] \Scrooge
            \addlegendentry [blue!90!white!90,every mark/.append style={solid,fill=\pgfplotsmarklistfill},mark=o] \ATA
            \addlegendentry [red,every mark/.append style={solid,fill=\pgfplotsmarklistfill},mark=halfdiamond*] \OTO
            \addlegendentry [green!50!black!70,every mark/.append style={solid,fill=\pgfplotsmarklistfill},mark=oplus] \OTF
            \addlegendentry [orange,every mark/.append style={solid,fill=\pgfplotsmarklistfill},mark=triangle] \LL
            \addlegendentry [orange!30!black!80,every mark/.append style={solid,fill=\pgfplotsmarklistfill},mark=square] \Kafka
        \end{axis}
    \end{tikzpicture}
}

\newcommand{\graphGEO}{
    \begin{tikzpicture}[plot]
        \begin{axis}[
                title={(ii) Geographical \RSM{s}},
                scaled y ticks = false,
                xlabel={\axisReps},
                ylabel={\axisGeoTP},
                ymin=0,
                ymode=log,
                yticklabels={0,100,1000},
                xtick={4,10,19},
                legend to name={mainlegendGEO},
                legend columns=-1,
            ]

            \addplot [black,every mark/.append style={solid,fill=\pgfplotsmarklistfill},mark=pentagon*] table[x={X},y={Picsou}] {\graphGeo};
            \addplot [blue!90!white!90,every mark/.append style={solid,fill=\pgfplotsmarklistfill},mark=o] table[x={X},y={ATA}] {\graphGeo};
            \addplot [red,every mark/.append style={solid,fill=\pgfplotsmarklistfill},mark=halfdiamond*] table[x={X},y={OTO}] {\graphGeo};
            \addplot [green!50!black!70,every mark/.append style={solid,fill=\pgfplotsmarklistfill},mark=oplus] table[x={X},y={OTU}] {\graphGeo};
            \addplot [orange,every mark/.append style={solid,fill=\pgfplotsmarklistfill},mark=triangle] table[x={X},y={LL}] {\graphGeo};
            
            \addlegendentry [black,every mark/.append style={solid,fill=\pgfplotsmarklistfill},mark=pentagon*] \Scrooge
            \addlegendentry [blue!90!white!90,every mark/.append style={solid,fill=\pgfplotsmarklistfill},mark=o] \ATA
            \addlegendentry [red,every mark/.append style={solid,fill=\pgfplotsmarklistfill},mark=halfdiamond*] \OTO
            \addlegendentry [green!50!black!70,every mark/.append style={solid,fill=\pgfplotsmarklistfill},mark=oplus] \OTF
            \addlegendentry [orange,every mark/.append style={solid,fill=\pgfplotsmarklistfill},mark=triangle] \LL
        \end{axis}
    \end{tikzpicture}
}

\newcommand{\graphFStake}{
    \begin{tikzpicture}[plot]
        \begin{axis}[
                title={(i) Varying Stake},
                scaled y ticks = false,
                xlabel={\axisReps},
                ylabel={\axisGeoTP},
                ymin=0,
                xtick={4,7,10,13,16,19},
                legend to name={mainlegend2},
                legend columns=4,
            ]

            \addplot [orange,every mark/.append style={solid,fill=\pgfplotsmarklistfill},mark=asterisk] table[x={X},y={Picsou1}] {\graphStakeNewNoThrottle};
            \addplot [orange,every mark/.append style={solid,fill=\pgfplotsmarklistfill},mark=asterisk] table[x={X},y={Picsou1}] {\graphStakeNewThrottle};
            \addplot [green!50!black!70,every mark/.append style={solid,fill=\pgfplotsmarklistfill},mark=oplus] table[x={X},y={Picsou2}] {\graphStakeNewNoThrottle};
            \addplot [green!50!black!70,every mark/.append style={solid,fill=\pgfplotsmarklistfill},mark=oplus] table[x={X},y={Picsou2}] {\graphStakeNewThrottle};
            \addplot [cyan,every mark/.append style={solid,fill=\pgfplotsmarklistfill},mark=triangle] table[x={X},y={Picsou4}] {\graphStakeNewNoThrottle};
            \addplot [cyan,every mark/.append style={solid,fill=\pgfplotsmarklistfill},mark=triangle] table[x={X},y={Picsou4}] {\graphStakeNewThrottle};
            \addplot [red!50!black!70,every mark/.append style={solid,fill=\pgfplotsmarklistfill},mark=square] table[x={X},y={Picsou8}] {\graphStakeNewNoThrottle};
            \addplot [red!50!black!70,every mark/.append style={solid,fill=\pgfplotsmarklistfill},mark=square] table[x={X},y={Picsou8}] {\graphStakeNewThrottle};
            \addplot [blue!90!white!90,every mark/.append style={solid,fill=\pgfplotsmarklistfill},mark=o] table[x={X},y={Picsou16}] {\graphStakeNewNoThrottle};
            \addplot [blue!90!white!90,every mark/.append style={solid,fill=\pgfplotsmarklistfill},mark=o] table[x={X},y={Picsou16}] {\graphStakeNewThrottle};
            \addplot [orange!30!black!80,every mark/.append style={solid,fill=\pgfplotsmarklistfill},mark=diamond] table[x={X},y={Picsou32}] {\graphStakeNewNoThrottle};
            \addplot [orange!30!black!80,every mark/.append style={solid,fill=\pgfplotsmarklistfill},mark=diamond] table[x={X},y={Picsou32}] {\graphStakeNewThrottle};
            \addplot [orange!30!black!80,every mark/.append style={solid,fill=\pgfplotsmarklistfill},mark=diamond] table[x={X},y={Picsou64}] {\graphStakeNewNoThrottle};
            \addplot [orange!30!black!80,every mark/.append style={solid,fill=\pgfplotsmarklistfill},mark=diamond] table[x={X},y={Picsou64}] {\graphStakeNewNoThrottle};

            \addlegendentry [orange,every mark/.append style={solid,fill=\pgfplotsmarklistfill},mark=asterisk] {Picsou1}
            \addlegendentry [green!50!black!70,every mark/.append style={solid,fill=\pgfplotsmarklistfill},mark=oplus] {Picsou2}
            \addlegendentry [cyan!50!black!70,every mark/.append style={solid,fill=\pgfplotsmarklistfill},mark=triangle] {Picsou4}
            \addlegendentry [red!50!black!70,every mark/.append style={solid,fill=\pgfplotsmarklistfill},mark=square] {Picsou8}
            \addlegendentry [blue!90!white!90,every mark/.append style={solid,fill=\pgfplotsmarklistfill},mark=o] {Picsou16}
            \addlegendentry [orange!30!black!80,every mark/.append style={solid,fill=\pgfplotsmarklistfill},mark=diamond] {Picsou32}
            \addlegendentry [orange!30!black!80,every mark/.append style={solid,fill=\pgfplotsmarklistfill},mark=diamond] {Picsou64}
        \end{axis}
    \end{tikzpicture}
}

\newcommand{\graphDR}{
    \begin{tikzpicture}[plot]
        \begin{axis}[
                title={(i) Disaster Recovery},
                scaled y ticks = false,
                xlabel={\axisMsg},
                ylabel={Throughput (MB/s)},
                ymin=0,
                xtick={.24,.5,2.0,4.0,19},
                xticklabels={.24,.5,2.0,4.0,19},
                xmode=log,
                log ticks with fixed point,
                legend to name={mainlegendApps},
                legend columns=-1,
            ]
            \addplot [black,every mark/.append style={solid,fill=\pgfplotsmarklistfill},mark=pentagon*] table[x={X},y={Picsou}] {\graphDRPlot};
            \addplot [blue!90!white!90,every mark/.append style={solid,fill=\pgfplotsmarklistfill},mark=o] table[x={X},y={ATA}] {\graphDRPlot};
            \addplot [red,every mark/.append style={solid,fill=\pgfplotsmarklistfill},mark=halfdiamond*] table[x={X},y={OTO}] {\graphDRPlot};
            \addplot [green!50!black!70,every mark/.append style={solid,fill=\pgfplotsmarklistfill},mark=oplus] table[x={X},y={GeoBFT}] {\graphDRPlot};
            \addplot [orange,every mark/.append style={solid,fill=\pgfplotsmarklistfill},mark=triangle] table[x={X},y={LL}] {\graphDRPlot};
            \addplot [orange!30!black!80,every mark/.append style={solid,fill=\pgfplotsmarklistfill},mark=square] table[x={X},y={KFK}]{\graphDRPlot};
            \addplot [purple!50!black!50,every mark/.append style={solid,fill=\pgfplotsmarklistfill},mark=asterisk] table[x={X},y={ETCD}]{\graphDRPlot};

            \addlegendentry [black,every mark/.append style={solid,fill=\pgfplotsmarklistfill},mark=pentagon*] \Scrooge
            \addlegendentry [blue!90!white!90,every mark/.append style={solid,fill=\pgfplotsmarklistfill},mark=o] \ATA
            \addlegendentry [red,every mark/.append style={solid,fill=\pgfplotsmarklistfill},mark=halfdiamond*] \OTO
            \addlegendentry [green!50!black!70,every mark/.append style={solid,fill=\pgfplotsmarklistfill},mark=oplus] \OTF
            \addlegendentry [orange,every mark/.append style={solid,fill=\pgfplotsmarklistfill},mark=triangle] \LL
            \addlegendentry [orange!30!black!80,every mark/.append style={solid,fill=\pgfplotsmarklistfill},mark=square] \Kafka
            \addlegendentry 
            [purple!50!black!50,every mark/.append style={solid,fill=\pgfplotsmarklistfill},mark=asterisk] \ETCD
        \end{axis}
    \end{tikzpicture}
}

\newcommand{\graphCCF}{
    \begin{tikzpicture}[plot]
        \begin{axis}[
                title={(ii) Data Reconciliation},
                scaled y ticks = false,
                xlabel={\axisMsg},
                ylabel={Throughput (MB/s)},
                ymin=0,
                xtick={.24,.5,2.0,4.0,8.0,19},
                xticklabels={.24,.5,2.0,4.0,8.0,19},
                xmode=log,
                log ticks with fixed point,
                legend to name={mainlegendApps},
                legend columns=-1,
                legend entries={\Scrooge{},\ATA,\OTO,\OTU,\LL,\Kafka, ETCD}
            ]
            \addplot [black,every mark/.append style={solid,fill=\pgfplotsmarklistfill},mark=pentagon*] table[x={X},y={Picsou}] {\graphCCFPlot};
            \addplot [blue!90!white!90,every mark/.append style={solid,fill=\pgfplotsmarklistfill},mark=o] table[x={X},y={ATA}] {\graphCCFPlot};
            \addplot [red,every mark/.append style={solid,fill=\pgfplotsmarklistfill},mark=halfdiamond*] table[x={X},y={OTO}] {\graphCCFPlot};
            \addplot [green!50!black!70,every mark/.append style={solid,fill=\pgfplotsmarklistfill},mark=oplus] table[x={X},y={GeoBFT}] {\graphCCFPlot};
            \addplot [orange,every mark/.append style={solid,fill=\pgfplotsmarklistfill},mark=triangle] table[x={X},y={LL}] {\graphCCFPlot};
            \addplot [orange!30!black!80,every mark/.append style={solid,fill=\pgfplotsmarklistfill},mark=square] table[x={X},y={KFK}]{\graphCCFPlot};
            \addplot [purple!50!black!50,every mark/.append style={solid,fill=\pgfplotsmarklistfill},mark=asterisk] table[x={X},y={ETCD}]{\graphCCFPlot};

            \addlegendentry [black,every mark/.append style={solid,fill=\pgfplotsmarklistfill},mark=pentagon*] \Scrooge
            \addlegendentry [blue!90!white!90,every mark/.append style={solid,fill=\pgfplotsmarklistfill},mark=o] \ATA
            \addlegendentry [red,every mark/.append style={solid,fill=\pgfplotsmarklistfill},mark=halfdiamond*] \OTO
            \addlegendentry [green!50!black!70,every mark/.append style={solid,fill=\pgfplotsmarklistfill},mark=oplus] \OTF
            \addlegendentry [orange,every mark/.append style={solid,fill=\pgfplotsmarklistfill},mark=triangle] \LL
            \addlegendentry [orange!30!black!80,every mark/.append style={solid,fill=\pgfplotsmarklistfill},mark=square] \Kafka
            \addlegendentry 
            [purple!50!black!50,every mark/.append style={solid,fill=\pgfplotsmarklistfill},mark=asterisk] \ETCD
        \end{axis}
    \end{tikzpicture}
}

\newcommand{\graphFail}{
    \begin{tikzpicture}[plot]
        \begin{axis}[
                title={(i) Crash Failures},
                scaled y ticks = false,
                xlabel={\axisReps},
                ylabel={\axisGeoTP},
                ymin=0, ymax=1500,
                ytick={0,300,600,900,1200,1500},
                yticklabels={0,300,600,900,1200,1500},
                xtick={4,7,10,13,16,19},
                legend to name={mainlegendCrash},
                legend columns=-1,
                legend entries={\Scrooge, \ATA, \OTU, \LL, \Kafka}
            ]

            \addplot [black,every mark/.append style={solid,fill=\pgfplotsmarklistfill},mark=pentagon*] table[x={X},y={Picsou}] {\graphMaxFail};
            \addplot [blue!90!white!90,every mark/.append style={solid,fill=\pgfplotsmarklistfill},mark=o] table[x={X},y={ATA}] {\graphMaxFail};
            \addplot [green!50!black!70,every mark/.append style={solid,fill=\pgfplotsmarklistfill},mark=oplus] table[x={X},y={OTU}] {\graphMaxFail};
            \addplot [orange,every mark/.append style={solid,fill=\pgfplotsmarklistfill},mark=triangle] table[x={X},y={LL}] {\graphMaxFail};
            \addplot [orange!30!black!80,every mark/.append style={solid,fill=\pgfplotsmarklistfill},mark=square] table[x={X},y={KFK}]{\graphMaxFail};

            \addlegendentry [black,every mark/.append style={solid,fill=\pgfplotsmarklistfill},mark=pentagon*] \Scrooge
            \addlegendentry [blue!90!white!90,every mark/.append style={solid,fill=\pgfplotsmarklistfill},mark=o] \ATA
            \addlegendentry [green!50!black!70,every mark/.append style={solid,fill=\pgfplotsmarklistfill},mark=oplus] \OTF
            \addlegendentry [orange,every mark/.append style={solid,fill=\pgfplotsmarklistfill},mark=triangle] \LL
            \addlegendentry [orange!30!black!80,every mark/.append style={solid,fill=\pgfplotsmarklistfill},mark=square] \Kafka
        \end{axis}
    \end{tikzpicture}
}

\newcommand{\graphBPhiList}{
    \begin{tikzpicture}[plot]
        \begin{axis}[
                title={(ii) Failure $\phi$ List},
                scaled y ticks = false,
                xlabel={\axisReps},
                ylabel={\axisGeoTP},
                ymin=0, ymax=1200,
                ytick={0,300,600,900,1200},
                yticklabels={0,300,600,900,1200},
                xtick={4,7,10,13,16,19},
                legend to name={mainlegend5},
                legend columns=-1,
            ]

            \addplot [orange,every mark/.append style={solid,fill=\pgfplotsmarklistfill},mark=asterisk] table[x={X},y={P0}] {\graphByzPhiList};
            \addplot [green!50!black!70,every mark/.append style={solid,fill=\pgfplotsmarklistfill},mark=oplus] table[x={X},y={P64}] {\graphByzPhiList};
            \addplot [cyan!50!black!70,every mark/.append style={solid,fill=\pgfplotsmarklistfill},mark=triangle] table[x={X},y={P128}] {\graphByzPhiList};
            \addplot [red!50!black!70,every mark/.append style={solid,fill=\pgfplotsmarklistfill},mark=square] table[x={X},y={P192}] {\graphByzPhiList};
            \addplot [black!90!white!90,every mark/.append style={solid,fill=\pgfplotsmarklistfill},mark=pentagon] table[x={X},y={P256}] {\graphByzPhiList};

            \addlegendentry [orange,every mark/.append style={solid,fill=\pgfplotsmarklistfill},mark=asterisk] {$\phi$0}
            \addlegendentry [green!50!black!70,every mark/.append style={solid,fill=\pgfplotsmarklistfill},mark=oplus] {$\phi$64}
            \addlegendentry [cyan!50!black!70,every mark/.append style={solid,fill=\pgfplotsmarklistfill},mark=triangle] {$\phi$128}
            \addlegendentry [red!50!black!70,every mark/.append style={solid,fill=\pgfplotsmarklistfill},mark=square] {$\phi$192}
            \addlegendentry [black!90!white!90,every mark/.append style={solid,fill=\pgfplotsmarklistfill},mark=pentagon] {$\phi$256}
        \end{axis}
    \end{tikzpicture}
}

\newcommand{\graphBAcking}{
    \begin{tikzpicture}[plot]
        \begin{axis}[
                title={(iii) Byzantine Acking},
                scaled y ticks = false,
                xlabel={\axisReps},
                ylabel={\axisGeoTP},
                ytick={100,1000},
                yticklabels={100,1000},
                xtick={4,7,10,13,16,19},
                ymode=log,
                legend to name={mainlegend4},
                legend columns=-1,
            ]

            \addplot [red,every mark/.append style={solid,fill=\pgfplotsmarklistfill},mark=oplus] table[x={X},y={Picsou-99}] {\graphByzAck};
            \addplot [green,every mark/.append style={solid,fill=\pgfplotsmarklistfill},mark=pentagon*] table[x={X},y={Picsou-0}] {\graphByzAck};
            \addplot [orange,every mark/.append style={solid,fill=\pgfplotsmarklistfill},mark=triangle] table[x={X},y={Picsou-ack-delay}] {\graphByzAck};
            \addplot [blue!90!white!90,every mark/.append style={solid,fill=\pgfplotsmarklistfill},mark=o] table[x={X},y={ATA}] {\graphByzAck};

            \addlegendentry [red,every mark/.append style={solid,fill=\pgfplotsmarklistfill},mark=oplus] {\Scrooge-Inf}
            \addlegendentry [green,every mark/.append style={solid,fill=\pgfplotsmarklistfill},mark=pentagon*] {\Scrooge-0}
            \addlegendentry [orange,every mark/.append style={solid,fill=\pgfplotsmarklistfill},mark=triangle] {\Scrooge-Delay}
            \addlegendentry [blue!90!white!90,every mark/.append style={solid,fill=\pgfplotsmarklistfill},mark=o] \ATA
        \end{axis}
    \end{tikzpicture}
}

%% file: intro-new.tex
\section{Introduction}
\label{s:intro}

Many organizations today use replicated state machines (\RSM{}) underpinned by consensus protocols to provide reliability, fault isolation, and disaster recovery. This includes key-value stores~\cite{etcd-raft,tidb,cockroach}, cluster managers~\cite{kubernetes},  and microservices~\cite{paxoscom, azureservicefabric,trafalgar}.
These \RSM{s} frequently need to communicate 
with each other in an efficient and timely manner. Etcd~\cite{etcd-raft} to Etcd mirroring over Kafka, for instance, is a popular approach for disaster recovery across clusters~\cite{disaster-recovery-confluent}.
Similarly, autonomous organizations often run their replicated key-value store locally for ease of management, but share access with other entities.  For example, conversations with government agencies reveal that, 
for operational sovereignty, services cannot be managed across agency borders. 
Instead, any shared information must be communicated across \RSM{s} and explicitly reconciled~\cite{ccf}.
Furthermore, in the blockchain ecosystem, there is a growing push towards interoperability, which requires distinct \RSM{s} (blockchains) to communicate~\cite{blockchain-interoperability-survey,blockchain-interop-survey}.



These examples speak of a common need: RSMs must support the ability to efficiently and reliably exchange messages with other RSMs \revision{that may or may not implement the same consensus protocol internally}{(whether they implement the same consensus protocol internally or not)}.

Unfortunately, existing solutions are either ad-hoc, offer vague \revision{or evolving}{(and evolving)} guarantees~\cite{blockchain-interop-survey}, 
rely on a trusted third-party~\cite{trustboost}), or 
require an expensive all-to-all broadcast~\cite{spanner,sharper}. For instance, Apache Kafka, the most popular approach for exchanging data across organizations, internally relies on a third RSM for safely sharing state.

All-to-all broadcast is even more problematic: while RSMs usually run within the same datacenter, there exist many RSMs which are geographically distributed. In these cases, cross-RSM communication will take place over WAN, which offers significantly reduced bandwidth \revision{at a much}{, and at much} higher dollar cost.
\revision{This frequently causes communication to become a bottleneck.}{}
\raf{this is awkward, but  I want to highlight that current applications are bottlenecked by cross-RSM communication} 
\sg{I understand what Reggie is saying, this can feel a bit too strong. Technically, all the solutions, we tested are bootlenecked by the communication but blockchain bridges, for instance, are compute bounded.}
\nc{Softened it}


Any system that allows RSMs to communicate should satisfy four requirements: 1) \textit{strong guarantees}: there should be a \revision{precise}{clean} 
and formal way to discuss RSM-RSM communication 2) \textit{robustness under failures}: actively malicious or crashed nodes should neither affect correctness nor cause system throughput to drop~\cite{aardvark}
3) \textit{low-overhead in the common-case}: for efficiency, an RSM to RSM communication protocol should send a single message with constant metadata in the failure-free case
4) \textit{generality}: arbitrary RSMs with heterogeneous sizes, communication, and fault models should be able to communicate. It should, for instance, be possible to link a Byzantine Fault Tolerant (\BFT) consensus protocol with a Crash Fault Tolerant (CFT) consensus algorithm 

To this effect, we first propose a new primitive, \CCCFull{} (\CCC{})\revision{, which}{that} 
can be used by two arbitrary RSMs to communicate. \CCC{} generalizes Reliable Broadcast to guarantee that if RSM $A$ sends $m$, at least one correct replica in RSM $B$ should receive $m$. 

We then introduce \Scrooge{}, a practical \CCC{} protocol that allows arbitrary \RSM{s} with heterogeneous communication and failure models to communicate efficiently. Designing a \CCC{} protocol that provides good performance in the failure-free case is fairly simple \revision{as a simple leader-to-leader broadcast suffices}{(example, a leader-to-leader broadcast suffices)}. 
The challenges instead arise from designing an efficient protocol that \textit{remains robust to failures}~\cite{aardvark}. The key to \Scrooge{}'s good \revision{and}{but} robust performance lies in observing that the C3B problem shares similar goals to TCP~\cite{computer-networks-book}. TCP seeks to offer reliable, ordered delivery between two hosts in a way that dynamically reacts to congestion and anomalies in the network. To do so at low cost, TCP leverages full-duplex communication and cumulative acknowledgments (ACKs) to asynchronously detect when messages have been received. In contrast, repeated ACKs of the same message reveal message loss. \Scrooge{} takes inspiration from these techniques and modifies them to account for the differences between C3B and TCP: 
1) unlike TCP, which is exclusively designed for point-to-point messaging, \Scrooge{} must handle many-to-many communication and 
disseminate knowledge of failed/successful message deliveries across many nodes, 
2) \Scrooge{} must ensure that no Byzantine participant will violate correctness or cause excessive retransmissions. In contrast, TCP does not consider malicious failures.

 
 To \revision{address these challenges}{achieve this}, \Scrooge{} introduces the notion of \quack{}s. A \quack{} is a cumulative quorum acknowledgment for a message $m$. It concisely communicates the fact that all messages up to $m$ have been reliably received by at least one correct node;
\textit{repeated} \quack{}s for $m$ indicate that the next message in the sequence was not received at the receiving RSM. 

Using \quack{}s in \Scrooge{} yields multiple benefits. First, it ensures \textit{generality}. It allows for \Scrooge{} to seamlessly work for crash fault tolerant systems as well as for both traditional and stake-based Byzantine fault tolerant protocols. \Scrooge{} makes no synchrony or partial synchrony assumption. 
Second, 
\Scrooge{}'s \quack{}-driven implementation has low overhead. In the failure-free case, \Scrooge{} sends each message only once, and requires only two additional counters per message. When failures do arise, \Scrooge{} remains \textit{robust} as its resend strategy minimizes the number of messages resent: no Byzantine node can unilaterally cause spurious message retransmissions.

%
Our results confirm \Scrooge{}'s strong guarantees. 
\Scrooge{} allows disparate protocols such as \pbft~\cite{pbftj}, Raft~\cite{raft} and Algorand~\cite{algorand} to communicate. 
On two real-world applications, Etcd Disaster Recovery~\cite{disaster-recovery-confluent} and a data reconciliation application~\cite{ccf}, \Scrooge{} achieves $2\times$ better performance than Kafka. In our microbenchmarks, when consensus is not the bottleneck, \Scrooge{} achieves $3.2\times$ better performance than a traditional All-to-All broadcast for small networks ($4$ nodes), and up to $24\times$ for large networks (19 nodes). In summary, this paper makes the following contributions:
\begin{enumerate}
    \item We introduce \CCCFull{} primitive, which allows for two RSMs to communicate robustly and efficiently.
    \item We present \Scrooge{}, a practical \CCC{} protocol. Key to \Scrooge's good performance is the use of \quack{s} (cumulative quorum acknowledgments), which precisely \revision{determines}{determine} when messages have definitely been
    received, or likely lost.
    \item \revision{We evaluate \Scrooge{} on realistic workloads, showing that it can successfully allow disparate RSMs to communicate more effectively that prior solutions.}{}
\end{enumerate}

%% file: prelim.tex
\section{Formalising the \CCC{} primitive }
\label{s:prelim}
\revision{We first introduce and formalize the \CCC{} primitive. \CCC{} is the blueprint for any communication protocol between RSMs and}{The \CCC{} primitive} should be sufficiently general to support various communication and failure models. 

\subsection{System Model}

\revision{We first discuss the necessary formalism.}{}
Consider a pair of communicating RSMS. For the sake of exposition,  we denote the sending \RSM{} as $\SMR{s}$ and the \revision{receiving \RSM{}}{receiver} as $\SMR{r}$.  In practice, communication between these \RSM{}s is full-duplex: both RSMs can send and receive messages.

Most modern \RSM{s} are either \textit{crash fault tolerant} (they guarantee consensus when up to $\f{}$ nodes crash) or \textit{Byzantine fault tolerant} (they guarantee consensus when up to $\f{}$ nodes behave arbitrarily). In line with \Scrooge's stated generality and efficiency goals, we adopt the \textit{UpRight} failure model~\cite{upright}. It allows us to consider Byzantine nodes and crashed nodes in a unified model, letting us design a system that optimizes for each type of failure. In the UpRight failure model, Byzantine nodes may exhibit {\em commission failures}; they may deviate from the protocol. All other faulty nodes \revision{may}{} suffer from \textit{omission failures} only: they follow the protocol but may fail to send/receive messages. Crashed nodes, for instance, suffer from permanent omission failures once crashed. Correct nodes, by definition, never fail%
\footnote{This is of course a simplification, necessary to say anything about an RSM. All practical systems assume that this $\f{}$ stays true at any given point in time, and that the set of failed nodes only changes. It allows making statements such as ``$\f{}+1$ votes ensure that at least one correct node participated''.}. 
In this setup, each \RSM{} consists of $\n{}$ replicas.  We denote the $j$-th replica at 
the $i$-th \RSM{} as $\Replica{i}{j}$ (where $i$ is either the sender or the receiver RSM). Each \RSM{} interacts with a set of clients, of which arbitrarily many can be faulty.

We say that an \RSM{} is safe despite up to $\rf{}$ commission failures and live despite up to $\uf{}$ failures of any kind.
For example, using the UpRight model, we can describe traditional \BFT{} and \CFT{} \RSM{s} using just one equation: $2\uf{}+\rf{}+1$;
Setting $\uf{}=\rf{}=\f{}$ yields a $3\f{}+1$ \BFT{} RSM and setting $\rf{}=0$ yields a $2\f{}+1$ \CFT{} \RSM{}.
Safety and liveness of any \RSM{} are defined as follows:
\begin{description}
\item[\bf Safety.]
If two correct replicas $\Replica{i}{1}$ and $\Replica{i}{2}$ \revision{in \RSM{} $\SMR{i}$}{} commit transactions $\Transaction{}$ and $\Transaction{'}$
at sequence number $k$, then $\Transaction{} = \Transaction{'}$.

\item[\bf Liveness.]
If a client sends a transaction $\Transaction{}$\revision{ to \RSM{} $\SMR{i}$}{}, \revision{correct replicas in $\SMR{i}$ will eventually commit $\Transaction{}$.}{then it will eventually receive a response for $\Transaction{}$.}
\end{description}

Note that we make no assumptions about the communication model of the underlying \RSM{}. 
We only assume messages are eventually delivered and that the \revision{receiving}{receiver} \RSM{} $\SMR{r}$ can verify whether a transaction was in fact 
committed by the sender \RSM{} $\SMR{s}$
(more details in Section~\ref{s:design}).

We generalize our system model to support \textit{shares}. The notion of share is used in stake-based \BFT{} consensus protocols where the value of a share determines the \textit{decision-making power of a replica}~\cite{blockchain-book}. \revision{For a replica $\Replica{i}{j}$ in \RSM{} $\SMR{i}$, we write
$\share{ij}$ to represent the replica's stake}{We write
$\share{ij}$ for the share of $\Replica{i}{j}$}; the total amount of share in \RSM{} $\SMR{i}$ is then 
\revision{$\sum_{l=1}^{\abs{\n{i}}} \share{il}$}{$\sum_{l=1}^{\abs{\n{i}}} \share{l}$}.
A stake-based consensus algorithm is safe as long as replicas totaling no more than $\rf{i}$ shares \revision{deviate from the protocol}{behave maliciously}. The system is live as long as no more than $\uf{i}$ shares fail\revision{ to send/receive messages}.  Traditional \CFT{} and \BFT{} algorithms simply set all shares equal to one.

{\bf Adversary Model.}
We assume the existence of a standard adversary which can corrupt arbitrary nodes, delay and reorder messages but cannot break cryptographic primitives.
As is standard, we assume that reconfigurations are possible~\cite{ethereum-stake-withdrawl,algorand-stake-withdrawl} and
 that there exists a mechanism for an \RSM{} to reliably learn of the new configuration and/or share assignments. \revision{We provide more detail on how such a mechanism can be implemented in Section 4.4}{}

%% file: primitive.tex
\subsection{\CCCFull{}}
\label{ss:c3b}
The \CCC{} primitive enables efficient and reliable communication 
between a sender \RSM{} and a receiver \RSM{}. 

%
To formalize \CCC{}, we first need to define two new communication primitives that express exchanging messages between RSMs. These operations are at a coarser granularity than the traditional \textit{send} and \textit{receive} operations, which define the operations performed by a specific node or replica.
\vspace{-6pt}
\begin{description}
\item[\bf Transmit.]
If a correct replica in \revision{$\SMR{s}$}{$\SMR{r}$} invokes \CCC{} on message $m$, 
we say that \RSM{} $\SMR{s}$ {\em transmits} message $m$ to \RSM{} $\SMR{r}$.
\item[\bf Deliver.]
If a correct replica from $\SMR{r}$ outputs message $m$, we say that \RSM{} $\SMR{r}$ {\em delivers} message $m$ from $\SMR{s}$.
\end{description}
\vspace{-2pt}
We can then define the two correctness properties that all \CCC{} implementations must satisfy:
\vspace{-6pt}
\begin{description}
\item[\bf Eventual Delivery.]
If \RSM{} $\SMR{s}$ transmits message $m$ to \RSM{} $\SMR{r}$, then $\SMR{r}$ will eventually deliver $m$.
\item[\bf Integrity.]
For every message $m$, an \RSM{} $\SMR{r}$ delivers $m$ from $\SMR{s}$ if and only if $\SMR{s}$
transmitted $m$ to $\SMR{r}$.
\end{description}
\vspace{-2pt}
\revision{Note}{In addition, note} that the deliver operation requires only that one correct node receives the message, not that all correct nodes receive it. This is by design: our goal is to make the \CCC{} condition as flexible as possible to suit application needs. In practice, it is easy to strengthen this condition to \revision{either}{} guarantee delivery to all nodes \revision{or to establish ordering between \RSM{}s}{}, as correct nodes \revision{in the receiving \RSM{} can simply broadcast or invoke consensus on delivered messages}{ can simply broadcast the message to all other correct nodes}.\revision{}{ \Scrooge{}, for instance, naturally ensures all correct nodes deliver the message. }
\revision{In}{
Also in} the name of generality, the \CCC{} primitive talks about a single message only, and does not worry about ordering across messages. \Scrooge{}, nonetheless, uses knowledge of ordering for efficiency, as we describe next.


%% file: algo-nc.tex
\section{Design Overview}
\label{s:design}

\Scrooge{} \textit{efficiently} implements the \CCC{} primitive while remaining \textit{general}  and \textit{robust} under failures. The design of \Scrooge{} is centered around three pillars:
\begin{enumerate}[wide,nosep,label=(P\arabic*),ref={P\arabic*}]

\item \label{g:zero} {\bf Efficiency.} In the (common) failure-free \revision{case, where messages are received in a timely fashion}{and synchronous case}, \Scrooge{} should only send a single copy of \revision{each}{the} message across \RSM{}s, and no more than $O(n)$ copies \revision{within a }{intra-}cluster.  Any additional metadata sent as part of \Scrooge{} should have constant size.

\item \label{g:hetero} {\bf Generality.} \Scrooge{} \revision{should}{must} support \RSM{s} of arbitrary sizes, with diverse failure models and communication models, including crash and Byzantine faults as well as synchronous and asynchronous networks. The protocol logic must also work well for both traditional \BFT{} systems and newer Proof-of-Stake protocols, where a replica's share \revision{determines}{illustrates} the weight that its vote carries.

\item \label{g:failures} {\bf Robustness.} \Scrooge{} \revision{should}{must} remain robust to failures. Crashed or malicious replicas should have minimal impact on performance~\cite{aardvark}. 
There is a tension here: while the protocol should aggressively resend dropped messages to minimize latency, Byzantine nodes should not cause correct nodes to resend messages, which can spuriously hurt throughput.
 
\end{enumerate}

Much like TCP flows, communication between RSMs is \textit{streaming}, \textit{long-running} and often \textit{full-duplex}. \Scrooge{} thus draws inspiration from TCP's approach to congestion control and message loss to guarantee efficiency and robustness. 


Two ideas are central to TCP's good performance: 1) leveraging full-duplex communication and 2) cumulative ACKing. In TCP, nodes simultaneously exchange messages, and TCP leverages this bidirectional communication to piggyback acknowledgments \revision{onto}{on the} messages and minimize bandwidth requirements. Cumulative ACKing  then keeps these acknowledgments small: with a single counter $k$, a receiver informs a sender that it has received all packets with sequence number up to $k$. 
Receiving \revision{a repeated counter with value}{repeated counter values} $k$ instead informs the sender that the packet with sequence number $k+1$ is \revision{either}{likely} lost or delayed.

For \textit{efficiency}, \Scrooge{} also leverages full-duplex communication and cumulative ACKing. However, \revision{}{unlike TCP,} \Scrooge{} must also handle many-to-many communication as each RSM consists of multiple replicas. 
For \textit{generality}, \Scrooge{} must support crashed nodes, Byzantine nodes, as well as Byzantine nodes whose stake determines their voting power;  
\Scrooge{} makes use of the \textit{UpRight} \revision{fault}{failure} model to simultaneously handle crash and Byzantine nodes, and leverages the mathematics of \textit{apportionment} to work seamlessly with staked-based systems. For \textit{robustness}, \Scrooge{} must ensure that replicas cannot trigger spurious message retransmissions. \Scrooge{} uses \quack{}s to determine when a message has definitely been received or is likely lost.


\par \textbf{Overview.}  \Scrooge{}'s protocol logic can be  divided into the following logical components.
\par (1) \textit{Consensus}:  On either side of \Scrooge{} lies a replicated state machine.  Each \RSM{} receives requests from clients and runs a consensus protocol that commit these requests on each \RSM{} replica.
\Scrooge{} assumes two properties of consensus to guarantee correctness: \revision{first,}{} all replicas eventually receive all messages, and \revision{second,}{} all replicas agree on the content of each slot in the log.
\par (2) \textit{Invoking \Scrooge{}}: \revision{If an \RSM{} wants to transmit a request,}{} each replica forwards the committed request to the co-located \Scrooge{} library. \revision{\RSM{}s}{Replicas} are not required to forward every committed message to \Scrooge{}. This step is application-specific. For example, if two organizations only share a subset of their data, the \RSM{} will only transmit messages that touch these particular objects.
\par (3) \textit{Transmitting a message}: \Scrooge{} sends the message on behalf of the sending \RSM{}. In line with our stated efficiency goals, \Scrooge{} ensures that, in the absence of failures and during periods of synchrony, a \textit{single sender} forwards each request to a \textit{single replica} in the receiving \RSM{}. To minimize the risk of repeated failures caused by a faulty sender or receiver, \Scrooge{} carefully rotates sender-receiver pairs. In doing so, it ensures that every sender will eventually communicate with a correct receiver and vice-versa. 
\par (4) \textit{Detecting successful or failed sends:} 
 \Scrooge{} must quickly determine whether a message has definitely been received (and thus can be garbage collected) or has definitely been dropped or delayed (distinguishing between message drops and delays is not possible in an asynchronous system). 
Failure detection must be accurate to prevent \revision{Byzantine}{failed} nodes from causing spurious re-transmissions; it must be efficient and should not require additional communication between nodes.  \Scrooge{} adapts TCP's cumulative acknowledgment approach to detect when messages have been received or dropped, even \revision{when}{as} malicious replicas can lie. These acknowledgments are piggybacked on incoming messages, thus minimizing overhead. 
\par (5) \textit{Retransmissions}: When the protocol detects that a message has been dropped (with high confidence), \Scrooge{} intelligently chooses the node responsible for resending the message. Unfortunately, \revision{concurrent node failures can cause multiple messages to be dropped simultaneously}{Byzantine nodes could still selectively drop these messages such that throughput craters}. 
To address this issue, \Scrooge{}  includes (constant size) information about which messages have been lost, allowing the protocol to recover \revision{multiple}{} dropped messages in parallel. 
Traditional TCP can eschew this constraint as it assumes failures are rare and considers point-to-point communication only.

\par \textbf{Correctness} We defer a full description of correctness and proofs to Appendix~\ref{ss:correctness} and~\ref{subsec:bounds}.

\section{Protocol Design}

We now describe each component of the protocol: transmitting a message, detecting successful/failed sends, and retransmissions.  
For clarity of exposition, we describe the protocol as consisting of a sender RSM \revision{$\SMR{s}$}{} and a receiver RSM \revision{$\SMR{r}$}{} proceeding in synchronous timesteps. In practice, nodes operate independently and act as both the sender and the receiver.
We start by describing \Scrooge{}'s behavior in the common case  (\S\ref{ss:failurefree}) before considering failures (\S\ref{ss:failures}). 
We add support for stake in \S\ref{s:stake}. We include protocol pseudocode in Appendix~\ref{subsec:pseudocode}.

\begin{figure}[t]
    \centering
    \begin{subfigure}[t]{\columnwidth}
         \centering
         \includegraphics[width=\columnwidth]{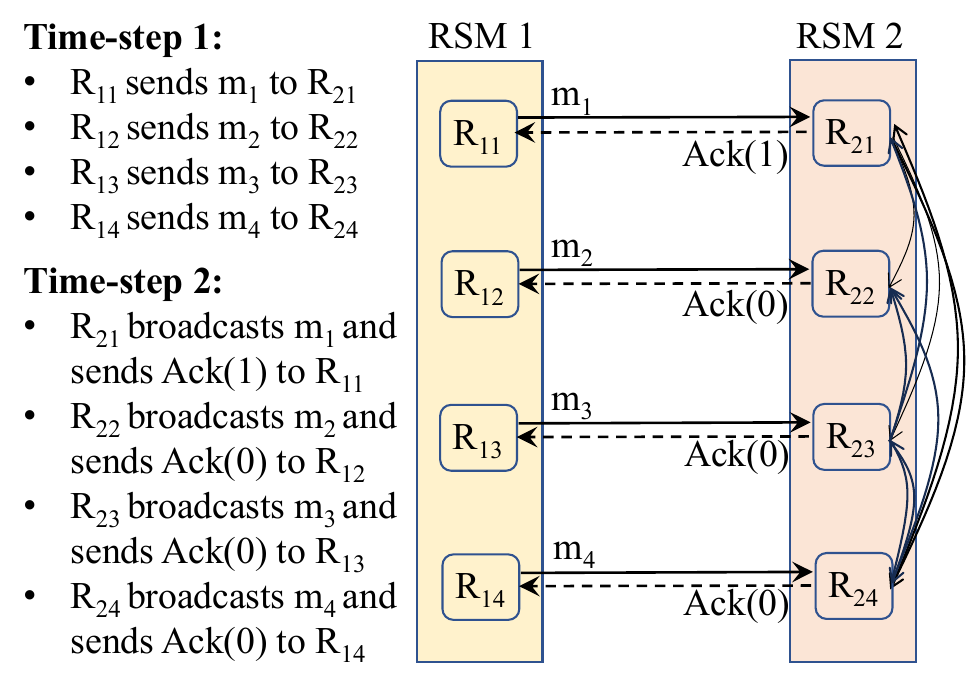}
         \label{sfig:first}
     \end{subfigure}
     \begin{subfigure}[t]{\columnwidth}
         \centering
         \includegraphics[width=\columnwidth]{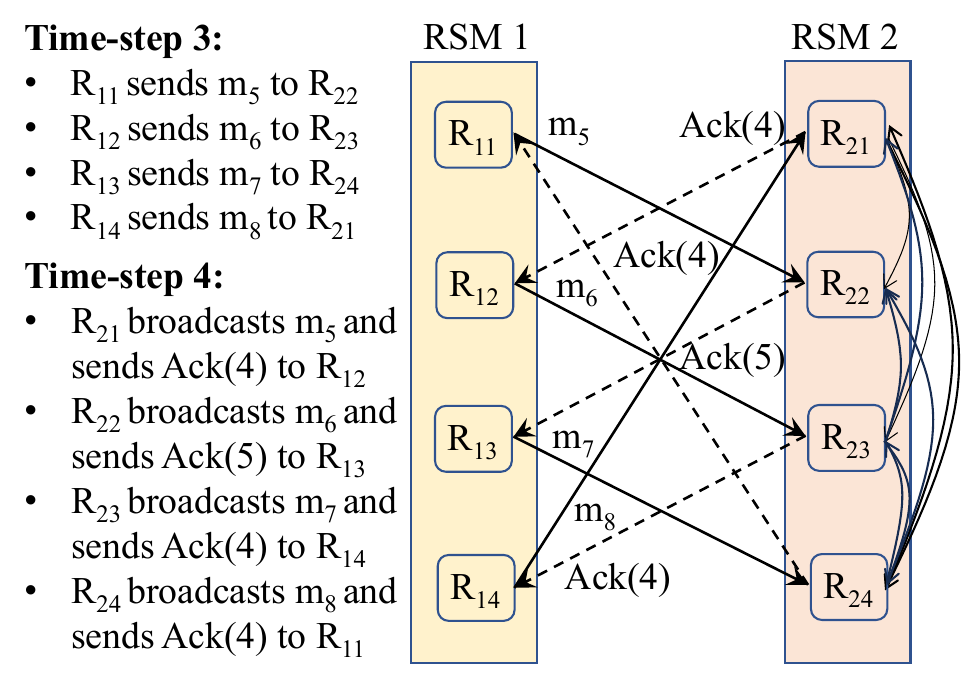}
         \label{sfig:second}
     \end{subfigure}%
    \vspace{-4mm}
    \caption
    {Example failure-free run in \Scrooge{}}
    \label{fig:round-robin}
\end{figure}

\subsection{Failure-free behavior}
\label{ss:failurefree}

\par \textbf{Sending a message.} \Scrooge{}'s send logic has three goals: 1) minimize the number of nodes sending the same message
2) maximize the chances that a message will be received quickly 3) asynchronously disseminate knowledge of received messages to other nodes. 
\Scrooge{} achieves these goals
by round-robin partitioning the set of requests across all sending \RSM ~replicas 
and rotating receiver nodes every round.

By definition, each replica in an \RSM{} contains a log of committed requests, a subset of which should be transmitted to the other \RSM{}. 
\revision{\Scrooge{}  assumes that each request transmitted through \Scrooge{} is of the form $\SignMessage{m, \Seqn, k^\prime}{\Qusign{s}}$, 
where $m$ is a request committed at sequence number $\Seqn$ by a quorum of replicas in \RSM{} $\SMR{s}$. Each protocol sets a specific threshold $t$ above which the request has acquired sufficiently many signatures
$\Qusign{s}$ to be deemed committed. $k^\prime$ is an optional sequence number that denotes the position of the message in the stream of messages that will be transmitted through \Scrooge{}. $k^\prime$ must be sequentially increasing; $k^\prime=\bot$ indicates that the message should not be transmitted. Including both $k$ and $k^\prime$ allows $\SMR{s}$ to filter which messages will be transmitted to $\SMR{r}$.
}{}

\Scrooge{} evenly partitions the task of transmitting \revision{committed messages}{a committed message} across all replicas such that each message is sent by a single node: \revision{replica $\Replica{s}{l}$}{ the $l$-th \Scrooge{} node} sends messages with sequence number $(\Seqn' \bmod \n{s} \equiv l)$. 
\revision{Additionally, each sender rotates receivers on every send: if \RSM{} $\SMR{r}$ has size $\n{r}$ and replica $\Replica{s}{l}$ last sent to replica $\Replica{r}{j}$, then $\Replica{s}{l}$ will next send to $\Replica{r}{J}$, where $J \equiv j + 1 \bmod \n{r}$.}{Each replica rotates destinations on every send: if the receiver \RSM{} has size $\n{r}$, and $j$ is the identifier of the previous recipient, then the $l$-th sender will send to receiver replica $(j+1) \bmod \n{r}$. }
Node IDs themselves are assigned by \Scrooge{} using a verifiable source of randomness~\cite{algorand} such that malicious nodes cannot choose specific positions in the rotation. Note that, as is standard in TCP, we allow senders to transmit a window of messages in parallel.
\par\textbf{Receiving 
a message} \revision{Upon receiving a message, the $j$-th replica $\Replica{r}{j}$ checks that the message $\SignMessage{m, \Seqn, k^\prime}{\Qusign{s}}$ is valid (the message has provably been committed by the sender \RSM{}) and if so, broadcasts it to the other nodes in its \RSM{}. Importantly, the receiving node does not need to recommit the message. It can simply apply it to its local state as mandated by the application logic.}

%
%

\begin{figure}[t]
    \centering
    \begin{subfigure}[b]{0.49\columnwidth}
         \centering
         \includegraphics[width=\textwidth]{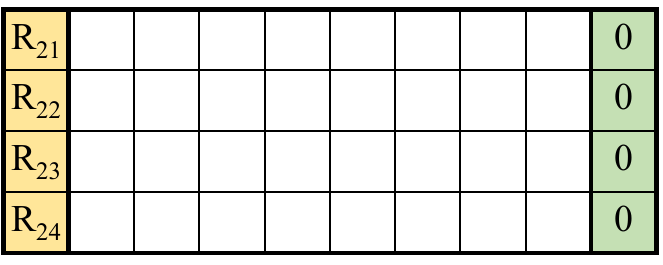}
         \caption{Time-step 1}
         \label{sfig:initial}
     \end{subfigure}
     \begin{subfigure}[b]{0.49\columnwidth}
         \centering
         \includegraphics[width=\textwidth]{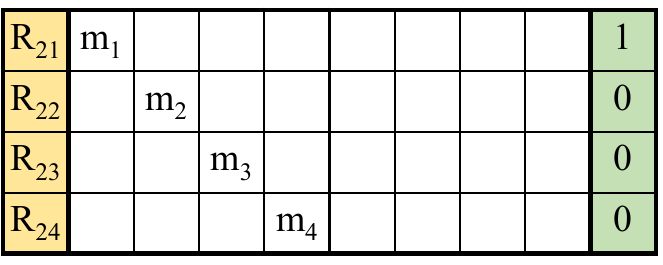}
         \caption{Time-step 2}
         \label{sfig:first-set}
     \end{subfigure}
     \begin{subfigure}[b]{0.49\columnwidth}
         \centering
         \includegraphics[width=\textwidth]{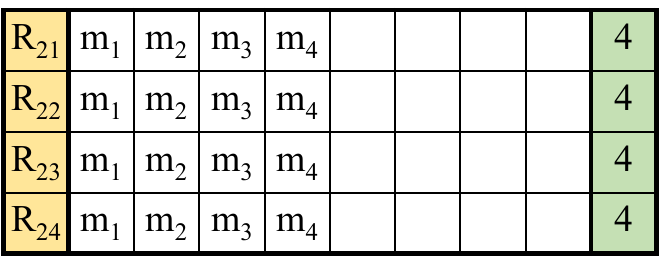}
         \caption{Time-step 3}
         \label{sfig:first-broadcast}
     \end{subfigure}
     \begin{subfigure}[b]{0.49\columnwidth}
         \centering
         \includegraphics[width=\textwidth]{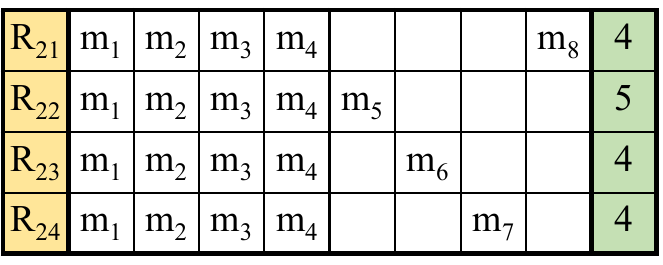}
         \caption{Time-step 4}
         \label{sfig:second-set}
     \end{subfigure}
     \begin{subfigure}[b]{0.49\columnwidth}
         \centering
         \includegraphics[width=\textwidth]{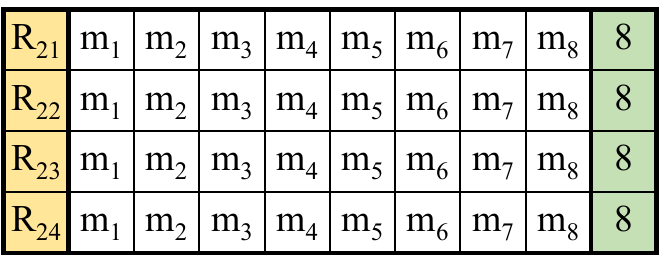}
         \caption{Time-step 5}
         \label{sfig:second-broadcast}
     \end{subfigure}
    \caption{Receiver's view of events in Figure~\ref{fig:round-robin} in time-steps.}
    \label{fig:ack-counter}
\end{figure}

Rotating sender-receiver pairs in this way guarantees that every pair of replicas will eventually exchange messages and ensures that (1) information about the state of each node is propagated to every other node in the system, and (2) no sender is continuously sending to a faulty replica (or vice-versa). This process is also essential to bounding the number of retransmissions needed with failures (\S\ref{ss:failures}).

We illustrate \Scrooge's logic in Figure~\ref{fig:round-robin}.
For clarity of exposition, we assume that 1) in each time-step, \revision{each}{a} replica completes all relevant tasks in parallel, 2) all \revision{sent}{} messages are received in the next time-step, and 3) only one \RSM{} sends, the other only acks. \revision{In our implementation, acks are piggy-backed on messages.}{}
Consider a system with $\n{s} = \n{r} = 4$ replicas ($\uf{}= \rf{} = 1$).
In time-step 1, replicas $\Replica{1}{1}$, $\Replica{1}{2}$, $\Replica{1}{3}$, and $\Replica{1}{4}$
of $\RSM{}_1$ send messages
 $m_1$, $m_2$, $m_3$, and $m_4$ to receivers $\Replica{2}{1}$, $\Replica{2}{2}$, $\Replica{2}{3}$, and $\Replica{2}{4}$, respectively.  
In time-step 2, these receivers internally broadcast these messages to the other nodes in their \RSM{}.
Concurrently, $\Replica{2}{1}$, $\Replica{2}{2}$, $\Replica{2}{3}$, and $\Replica{2}{4}$ 
acknowledge receipt of these messages and send $\Ack{1}$, $\Ack{0}$, $\Ack{0}$, and $\Ack{0}$ to senders 
$\Replica{1}{1}$, $\Replica{1}{2}$, $\Replica{1}{3}$, and $\Replica{1}{4}$.
We discuss acknowledgments later in the section.
In time-step 3, $\Replica{1}{1}$, $\Replica{1}{2}$, $\Replica{1}{3}$, and $\Replica{1}{4}$ rotate receivers and
send messages $m_5$, $m_6$, $m_7$, and $m_8$ to $\Replica{2}{2}$, $\Replica{2}{3}$, $\Replica{2}{4}$, and $\Replica{2}{1}$.
In time-step 4, receivers once again broadcast the received messages in their \RSM{}.
Simultaneously, $\Replica{2}{1}$, $\Replica{2}{2}$, $\Replica{2}{3}$, and $\Replica{2}{4}$ rotate receivers for their acknowledgements, 
and send $\Ack{4}$, $\Ack{5}$, $\Ack{4}$, and $\Ack{4}$ to senders 
$\Replica{1}{2}$, $\Replica{1}{3}$, $\Replica{1}{4}$, and $\Replica{1}{1}$. 

\begin{figure}[t]
    \centering
    \begin{subfigure}[b]{0.49\columnwidth}
         \centering
         \includegraphics[width=\textwidth]{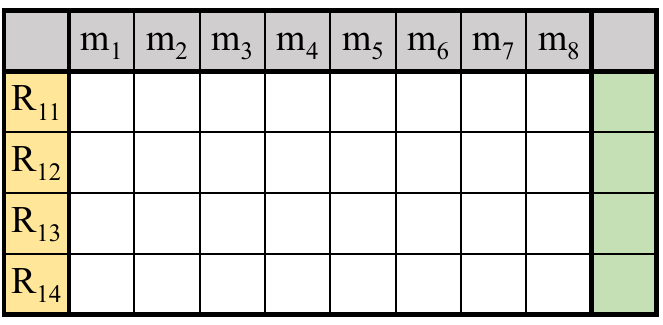}
         \caption{Time-step 1}
         \label{sfig:init-ack}
     \end{subfigure}
     \begin{subfigure}[b]{0.49\columnwidth}
         \centering
         \includegraphics[width=\textwidth]{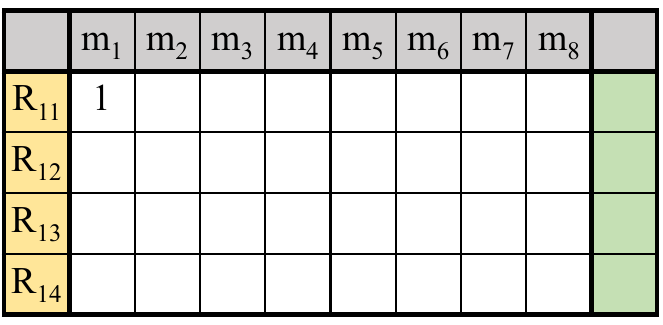}
         \caption{Time-step 3}
         \label{sfig:first-ack}
     \end{subfigure}
     \begin{subfigure}[b]{0.49\columnwidth}
         \centering
         \includegraphics[width=\textwidth]{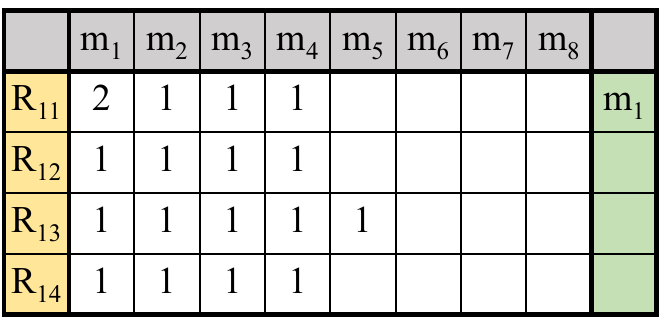}
         \caption{Time-step 5}
         \label{sfig:second-ack}
     \end{subfigure}
    \caption{Sender's view of events in Figure~\ref{fig:round-robin} in time-steps.}
    \label{fig:quack-counter}
\end{figure}

\par \textbf{Detecting successful sends.} To guarantee correctness, committed messages must eventually be received by a correct node in the receiving \RSM{}. Every node in the sending \RSM{} (not just the sender) must thus learn whether a message has \textit{definitely} been received by a correct node. This is necessary to preclude correct nodes from unnecessarily resending messages.
There are three primary challenges: (1) malicious nodes may lie about the set of messages received, (2) for efficiency, \Scrooge{} should not require nodes within an \RSM{} to exchange information beyond the necessary message broadcast, and (3) any additional metadata should be small. \Scrooge{} realizes these goals through \textit{cumulative quorum acknowledgments} (or \quack{}s). A cumulative quorum acknowledgment with value $\Seqn$ proves to the sending \RSM{} that all messages with sequence number up to $\Seqn$ have been received by at least one correct replica. 
More specifically, each replica, upon receiving a message with sequence number $\Seqn$, inserts it into a sorted list containing all previously received messages. The replica \revision{then identifies}{does not, however, \textit{selectively} acknowledge $m_k$. Instead, it identifies} the highest message $m_p$ in the list for which all messages with a smaller sequence number have been received. \revision{It then}{The replica} crafts an acknowledgment $\Ack{p}$ that cumulatively acknowledges receipt of messages $1$ to $p$. \Scrooge{} takes advantage of the full-duplex nature of the protocol to piggyback these acknowledgments onto the messages that the receiving \RSM{} is itself sending to the sending \RSM{}. If no such message exists, the \RSM{} sends a no-op.

On the sender side, each replica eventually receives messages and acknowledgments from all $\n{r}$ receiving replicas, thanks to \Scrooge's round-robin strategy.  Each replica maintains an $\n{r}$ sized array that summarizes the highest acknowledgment received from each replica of receiving \RSM{}.  A message $m_p$ is \quack{}ed if at least $\uf{r}+1$ replicas have acknowledged receipt of all messages up to $p$. As there are only $\uf{r}$ failed replicas\raf{We don't explicitly state that $u\ge r$}, one of these replicas must be correct. We thus have the guarantee that this correct replica will broadcast the message to all other remaining correct replicas. Note that \Scrooge{} additionally uses MACs \revision{to safely transmit \AckPlain{}s}{} when configured to handle Byzantine failures ($\rf{}>0$).
%
%

\par \textbf{Example.} Continuing with our example in Figure~\ref{fig:round-robin},
we highlight the protocol logic for both the receiver \RSM{} (Figure~\ref{fig:ack-counter})
and the sender \RSM{} (Figure~\ref{fig:quack-counter}).
Each row in Figure~\ref{fig:ack-counter}  describes a sorted list of messages received at each replica, 
with the last column denoting the highest cumulative acknowledgment for this node.
At time-step 1, all lists are empty and cumulative acknowledgment values are all set to $0$ (\ref{sfig:initial}). 
At time-step 2, 
receivers $\Replica{2}{1}$, $\Replica{2}{2}$, and $\Replica{2}{3}$ (\ref{sfig:first-set}) store messages $m_1$, $m_2$ and $m_3$.
$R_{21}$'s cumulative acknowledgment counter thus increases to 1, while others stay at 0 as they are still missing $m_1$. 
$\Replica{2}{1}$, $\Replica{2}{2}$, and $\Replica{2}{3}$ thus send $\Ack{1}$, $\Ack{0}$ and $\Ack{0}$ to the sender \RSM{}.
At time-step 3, each receiver, thanks to the internal broadcast mechanism, receives all four messages. All cumulative acknowledgment counters thus go to $4$ (\ref{sfig:first-broadcast}). 
By time-step 4, receivers $\Replica{2}{1}$, $\Replica{2}{2}$, and $\Replica{2}{3}$ have all received $m_8$, $m_5$ and $m_6$.
$\Replica{2}{2}$ has received messages $m_1$ to $m_5$, and thus updates its cumulative acknowledgment to $5$. In contrast,
\revision{}{through} $\Replica{2}{1}$ and $\Replica{2}{3}$ have received messages $m_8$ and $m_6$ respectively, they are missing $m_5$ and thus cannot
yet update their cumulative acknowledgment counter. 
$\Replica{2}{1}$, $\Replica{2}{2}$, and $\Replica{2}{3}$ send $\Ack{4}$, $\Ack{5}$ and $\Ack{4}$ back to the sending \RSM{}.
Finally, at time-step 5, the internal broadcast mechanism disseminates all these messages; each replica can update its cumulative acknowledgment to 8.

Now consider the sender-side logic (Figure~\ref{fig:quack-counter}), which processes these cumulative acknowledgments and determines when a \quack{}
has formed. 
Recall that a message is \quack{}ed at a replica if this replica receives $\uf{}+1=2$ acknowledgments for $m$.
Initially, all \quack{} counters  are empty (\ref{sfig:init-ack}). 
At time-step $3$ (\ref{sfig:first-ack}), $\Replica{1}{1}$
records that it has received an acknowledgment for $m_1$  ($\Ack{1}$) from $\Replica{2}{1}$. 
At time-step 5, (\ref{sfig:second-ack}), $\Replica{1}{1}$ receives $\Ack{4}$ from $\Replica{2}{4}$. 
It updates its local array, indicating that it has received acknowledgments for $m_1$ by two unique nodes, and marks $m_1$\ as \quack{}ed.
It also records the received acks for $m_2$, $m_3$, and $m_4$. 
Similarly, all other replicas indicate that they have received an ack for $m_1$, $m_2$, $m_3$, and $m_4$.
Additionally, $\Replica{1}{3}$ indicates it received an ack for $m_5$ as it received $\Ack{5}$. 

\par \textbf{Summary} The joint techniques of full-duplex communication, cumulative acking, and rotation of sender/receiver pairs allow \Scrooge{} to ensure that all \RSM{} replicas eventually learn which committed requests have been delivered. The protocol achieves this with only two additional counters and with no additional communication between replicas of an \RSM{} beyond the necessary broadcast.

\subsection{Handling Failures}
\label{ss:failures}
\revision{Faulty replicas can: 1) fail to send or broadcast messages sends and broadcasts, 2) send invalid messages to DDoS the network, 3) collude to repeatedly drop messages, and 4) send incorrect acknowledgments to break correctness or trigger spurious retransmissions.}{Crashed replicas can fail to complete a send or broadcast, while actively malicious replicas can 1) send invalid messages to DDOS the network, 2) collude to repeatedly drop messages be it by failing to send messages or failing to send acknowledgments, and 3) send incorrect acknowledgments to break correctness or trigger spurious retransmissions.} 
\Scrooge{} must effectively handle these failures without sacrificing correctness or performance. To this effect,  \Scrooge{} must quickly and reliably detect \textit{when} a message has \textit{definitely} been dropped or delayed and quickly retransmit it. The system must do so without any additional communication beyond resending the message itself.

\par \textbf{Detecting failed sends.} The protocol once again leverages \quack{}s to detect failed sends.   Recall that all sender replicas eventually obtain a \quack{} for every message that has \textit{definitely} been delivered. One can instead leverage duplicate 
\quack{}s to learn when a correct replica has \textit{not} received a specific message. 
In more detail, let us assume that a \quack{} for message $m_k$ has formed at $\Replica{s}{l}$. 
This \quack{} indicates that at least $\uf{} + 1$ (at least one correct) replicas have received every message up to message $m$ with sequence number $k$. 
A duplicate acknowledgment $\Ack{k}$ from one of these replicas implies that this replica claims not to have received the message at sequence number $k + 1$. Once a duplicate \quack{} forms for the $k$-th message at replica $\Replica{s}{l}$,  $\Replica{s}{l}$ learns that \revision{a correct replica is complaining about missing message $(k+1)$, and thus that the message has legitimately been lost or delayed}{, the $(k+1)$-th message has legitimately been lost or delayed (a correct replica is complaining about the missing message)}. All other replicas of the sending \RSM{} $\SMR{s}$ will eventually receive a duplicate \quack{} and thus detect the failed exchange.
The use of the UpRight failure model, which distinguishes actively malicious failures $\rf{}$ from all other failures, allows us to reduce the size of the duplicate \quack{}: while the initial \quack{} is of size    $\uf{} + 1$, duplicate \quack{}s must be of size $\rf{} + 1$ as they must be large enough to preclude actively malicious nodes from triggering spurious resends. In a system with only crash failures (when $\rf{}=0$), a single duplicate \Ack{} is sufficient to trigger a message resend: nodes may crash but do not lie. 

\par \textbf{Retransmitting the dropped message.} Upon detecting a failed send, the message must be quickly retransmitted. Just as a single replica was responsible for sending the initial message, \Scrooge{} ensures that a single replica is "elected" as the re-transmitter. It does so \textit{without} requiring additional communication between replicas. The protocol logic hinges on three observations: 1) all correct replicas know about all the messages that must be transmitted (by definition of an \RSM{}) and know who initially sent the message, 2) all correct replicas eventually learn about which messages have been \quack{}ed,
and 3) the number of repeated \quack{}s indicates the number of failed retransmissions. 
\Scrooge{} uses this information to compute the ID of the re-transmitter as: $sender_{new} = (sender_{original} + \#_{retransmit} )\bmod ~\n{s}$.  Each correct replica computes this function and retransmits the message if its ID matches $sender_{new}$.  Each retransmission round thus has a single sender.

\begin{figure}[t]
    \centering
    \begin{subfigure}[b]{0.49\columnwidth}
         \centering
         \includegraphics[width=\textwidth]{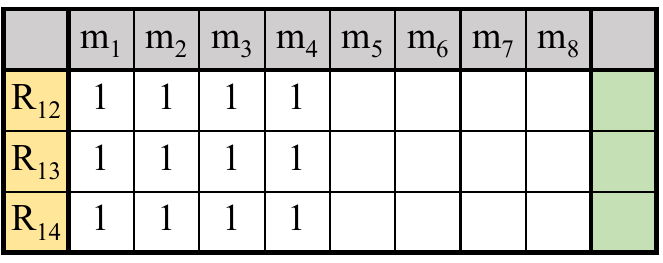}
         \caption{Time-step 5}
         \label{ssfig:init-ack}
     \end{subfigure}
     \begin{subfigure}[b]{0.49\columnwidth}
         \centering
         \includegraphics[width=\textwidth]{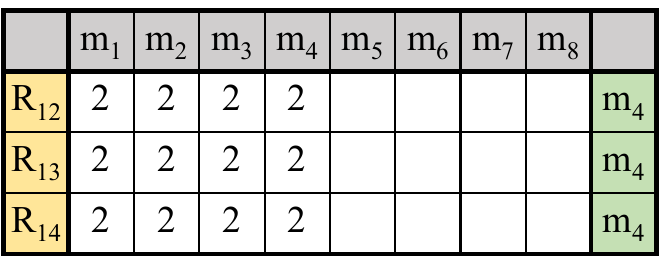}
         \caption{Time-step 7}
         \label{ssfig:first-ack}
     \end{subfigure}
     \begin{subfigure}[b]{0.49\columnwidth}
         \centering
         \includegraphics[width=\textwidth]{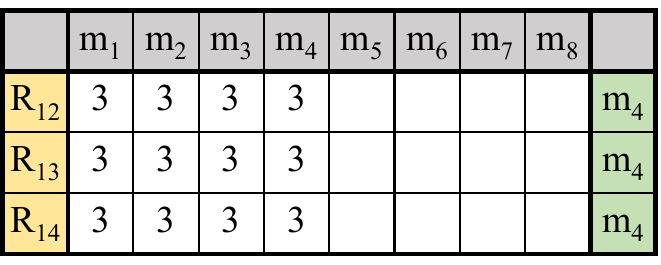}
         \caption{Time-step 9}
         \label{ssfig:second-ack}
     \end{subfigure}
     \begin{subfigure}[b]{0.49\columnwidth}
         \centering
         \includegraphics[width=\textwidth]{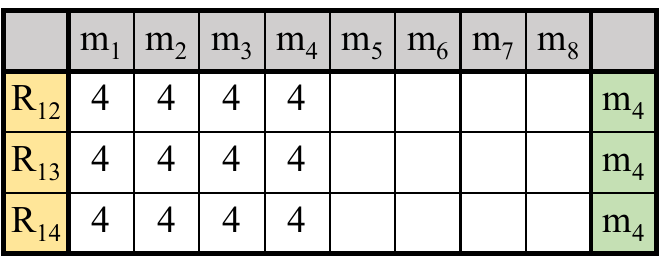}
         \caption{Time-step 11}
         \label{usfig:init-ack}
     \end{subfigure}
     \begin{subfigure}[b]{0.49\columnwidth}
         \centering
         \includegraphics[width=\textwidth]{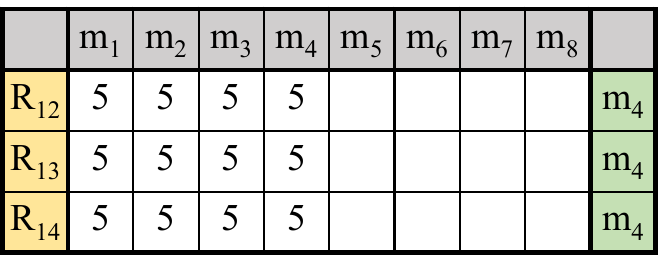}
         \caption{Time-step 13}
         \label{usfig:first-ack}
     \end{subfigure}
     \begin{subfigure}[b]{0.49\columnwidth}
         \centering
         \includegraphics[width=\textwidth]{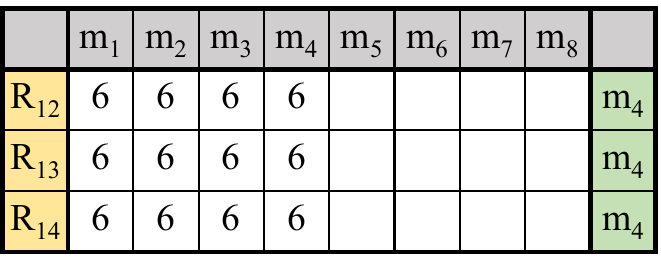}
         \caption{Time-step 15}
         \label{usfig:second-ack}
     \end{subfigure}
    \caption{Sender's view of events. $\Replica{1}{1}$ fails after TS 2  in Figure~\ref{fig:round-robin}. }
    \label{fig:fail-quack-counter}
\end{figure}

To illustrate, consider once again our initial example (Figure~\ref{fig:round-robin}), but this time, let us assume that sender replica $\Replica{1}{1}$ fails in time-step 2, after sending message $m_1$ but before sending messages $m_5$ and $m_9$. As a result, no receiver receives these messages. In Figure~\ref{fig:fail-quack-counter}, we time-step through this failure scenario. For simplicity, we assume that the receiving RSM sends periodic acks every time-step. 
%
As before, all non-failed replicas of $\RSM{}_1$ receive their first $\Ack{4}$ in time-step 5 (Figure~\ref{fig:round-robin});
In time-step 7, all replicas receive a second $\Ack{4}$ message, from a different node, allowing them to mark messages $m_1$ to $m_4$ as \quack{ed}. They continue receiving $\Ack{4}$ from \textit{distinct} replicas in time-step 9 and 11. Receivers cannot acknowledge any message greater than $m_4$ as they are yet to receive $m_5$.
In time-step 13, the senders receive their first duplicate $\Ack{4}$ message. 
By the end of time-step 15, the senders have received at least $\rf{r}+1 =2$ duplicate $\Ack{4}$ messages, confirming that $m_5$ is missing. $\Replica{1}{2}$ proceeds to resend $m_5$.

\par \textbf{The pitfalls of sequential recovery.} 
Unlike traditional TCP in which message drops are not adversarial, \revision{faulty}{Byzantine} replicas can carefully select which messages to drop. 
For instance,  in a $\n{} = 2\uf{}+ \rf{} +  1$ setup with $\uf{} = \rf{}= 1$, if a \revision{node omits}{malicious actor drops} all received messages, every fourth message will need to be resent.
In this setup, \Scrooge{} can \revision{hit}{introduce} a throughput bottleneck. 
A \quack{} conveys information about the \textit{lowest} message that has been dropped by the system, but says nothing about later messages.  This approach is optimal metadata-wise but serializes recovery: if messages $m_i$, $m_{i+4}$, $m_{i+8}$, etc. have all been dropped, 
resending $m_{i+8}$ first requires detecting the failed send of message $m_i$, retransmitting $m_i$, \quack{} $m_i$, before repeating
the same process for $m_{i+4}$. Only then can the failed send of $m_{i+8}$ be handled.

\par \textbf{Parallel Cumulative Acknowledgments.} 
To address this issue, we must augment our cumulative acknowledgments with a limited form of selective repeat~\cite{selective-repeat}.
Each receiver sends both a cumulative acknowledgment and a list \revision{summarizing the delivery status of up to $\phi$ messages past the sent cumulative acknowledgment}{($\phi$) of messages that it is missing}. The cumulative acknowledgment counter concisely summarizes the set of contiguous messages received so far. The $\phi$-list instead captures any "in-flight" missing messages. \revision{Sending $\phi$-lists over the network is efficient as the delivery status of each message takes at most one bit to encode. This list can further be reduced with techniques such as compression or bloom filters.}{}

Sender replicas can now, concurrently, form \quack{s} for $\phi$ concurrent messages and thus retransmit $\phi$ messages in parallel. \revision{This reduces latency without resorting to eager message resends}{}.
The maximum size of $\phi$-lists is an experiment-specific parameter.
The actual number of elements in a $\phi$-list depends on the number of \revision{in-flight}{missing} messages at the time of sending a cumulative acknowledgment.

%
%
\par \textbf{Analysis} During periods of synchrony (when messages are not \revision{dropped or delayed}{delayed} by the network), \Scrooge{} retransmits messages at most $\uf{s} + \uf{r}+ 1$ times. This limitation is fundamental to all \CCC{} protocols (Lemma 1 in Appendix~\ref{ss:correctness}). The number of resends can become a concern for latency if the number of failures is large. 
In practice, however, the probability of actually hitting this bound is small. \revision{}{In fact, assuming a fixed ratio of Byzantine nodes in the system, each with a random identifier (recall that \Scrooge{} assigns random IDs to nodes to prevent collusions), after 8 retries, the probability that a message was successfully delivered is already 99\%.}\revision{Intuitively, in a CFT or BFT system, each node is more likely than not to be correct. As such, the probability of continually selecting incorrect nodes in every sender-receiver pair decreases exponentially every retry. One can use this reasoning to provide strong bounds on the maximum number of retries when the network is well-behaved. We prove, for instance (in Appendix~\ref{subsec:bounds}) that \Scrooge{} needs to resend a message at most eight times to ensure that a message be delivered with 99\% probability, and at most 72 times to ensure a $100-10^{-9}\%$ success probability.}{even if there are far more than 8 failures across the two networks.} 

\revision{}{We prove (in Appendix~\ref{subsec:bounds}) that:

\begin{theorem}
 Given a sending \RSM{} with $\n{s}= \alp{s} \times \uf{s} + 1$ replicas and 
receiving \RSM{} with $\n{r}= \alp{r} \times \uf{r} + 1$ replicas,
\Scrooge{} needs to resend a message at most $72$ times to guarantee a $10^{-9}$ failure probability, irrespective of node count.
\end{theorem}
}

\subsection{Garbage Collection}
\label{ss:garbage}

At first glance, garbage collecting messages in \Scrooge{} appears straightforward. The sending \RSM{}, upon receiving a \quack{} for $m$, can garbage collect $m$ as the message has been received by a correct replica. Unfortunately, this approach can lead to scenarios in which \Scrooge{} stalls. Consider, for instance, an execution in which  sender $\Replica{s}{l}$ sends a message $m_\Seqn$ (at sequence number $\Seqn$) 
to replica $\Replica{r}{j}$ of \RSM{} $\SMR{r}$. Now, consider the case in which
$\Replica{r}{j}$ is \revision{faulty}{Byzantine} and broadcasts $m_\Seqn$ to precisely $\uf{r}+1$ replicas, $\uf{r}$ of which are faulty. These replicas reply to the sender RSM that $m$ has been successfully received, allowing for a \quack{} to form at the sender, and for message $m$ to be garbage collected. Unfortunately, if these $\uf{r}$ replicas then stop participating in the protocol, no \quack{} will ever form for any message with sequence number greater than $\Seqn$ (only one correct replica has seen $m$). Instead, the sending RSM will receive repeated duplicate acknowledgments for $m$, a message which it has already garbage collected!

We must consequently modify the garbage collection algorithm slightly. If a sending replica ever receives a duplicate \quack{} for message $m_{\Seqn'}$ where $\Seqn'<\Seqn$ \textit{after} having quacked and garbage collected message $m_\Seqn$, it includes, as additional metadata, the sequence number $\Seqn$ of its highest quacked message. This information conveys to the receiving RSM that all messages up until $\Seqn$ (included) have been received by \textit{some} correct node in the receiving RSM, but not necessarily the same one. Replicas in the receiving RSM, after having received $\rf{s}+1$ such messages (ensuring that at least one correct node is in the set), can then either (1) advance their cumulative acknowledgment counter to $\Seqn$ and mark message $m$ as received, or (2) obtain $m$ from other replicas in the \RSM{}. Only then can $m$ be garbage collected. We offer both strategies in \Scrooge{}. \raf{this paragraph/protocol can be simplified if we discuss the local quack behavior.}

\subsection{Reconfiguration}
 \revision{
 \Scrooge{} assumes that reconfigurations are possible but rare. It assumes that there exists a service indicating the set of nodes associated with each configuration. This is standard practice in the literature~\cite{etcd-raft,ethereum-stake-withdrawl,algorand-stake-withdrawl}.
 Knowledge of membership is either maintained internally in the RSM~\cite{moderately-complex-paxos}  or using an external configuration service~\cite{moderately-complex-paxos,etcd-raft}. Most existing state-of-the-art blockchain systems~\cite{ether,sui,algorand,hyperledger-fabric,MyTumbler,Aptos,ccf,MicrosoftACL,IBM-supply-chain-blockchain} also require known node membership. To deal with churn and scale, these systems work in epochs where it is assumed both node membership and relative stake are both publicly known and fixed. \Scrooge{} piggybacks on this assumption.}{
 \Scrooge{} assumes that reconfigurations are possible, but rare, and that there exists a service indicating the set of nodes associated with each configuration. This is standard practice in the literature~\cite{etcd-raft,ethereum-stake-withdrawl,algorand-stake-withdrawl}.
 }
  \Scrooge{} then only needs to ensure that the set of ACKs received for a particular message all match the same view and that the relevant ($\uf{}+1 / \rf{}+1$) threshold has been reached (for that view).
  
\revision{Messages acknowledged as delivered before a reconfiguration occurs do not need to be resent. Reconfiguration in an RSM, by definition, preserves any state across configurations. Messages not acknowledged as delivered before the reconfiguration begins must be resent as they may or may not have persisted.  After reconfiguration completes, \Scrooge{} simply resends messages for which it did not receive a quorum of acknowledgments in the prior configuration. 
}{}

\section{Weighted \RSM{s} -- \revision{Stake}{Stakes}}
\label{s:stake}

The current description of the protocol assumes that replicas have equal weight in the system.
When considering proof-of-stake systems like Algorand, each replica can instead hold differing \revision{amounts}{amount} of \revision{\textit{stake}}{\textit{stakes}} or \textit{shares} in the system. We write
$\share{ij}$ for the share of $\Replica{i}{j}$; the total amount of \revision{stake}{share} in \RSM{} $\SMR{i}$ is then
\revision{$\pmb{\Delta_{i}} = \sum_{l=1}^{\n{i}} \share{il}$}{$\n{i} = \sum_{l=1}^{\abs{\n{i}}} \share{l}$}. The \RSM{} is safe as long as replicas totalling no more than $\rf{i}$ shares \revision{deviate from the protocol}{behave maliciously}; the \RSM{} is live as long as replicas totalling no more than $\uf{i}$ shares \revision{omit messages}{fail}. The existence of stakes changes: (1) when a replica can establish a \quack{}, and (2) to whom a particular message must be sent. 

\subsection{Weighted \quack{}} It is straightforward to modify \quack{}s to deal with \revision{stake}{stakes}. Each cumulative acknowledgment message simply becomes weighted. 
The acknowledgment message from \revision{replica $\Replica{il}{}$}{an $l$-th replica} with share \revision{$\share{il}$}{$\share{l}$} has a weight \revision{$\share{il}$}{$\share{l}$} and  a \quack{} forms for message $m$ when 
the total weight of \revision{the}{} cumulative \quack{} for $m$ from \RSM{} $\SMR{i}$ is equal to $\uf{i}+1$.

\subsection{Sending a message} Identifying the appropriate sender-receiver pair for sending a message requires more care.  Traditional BFT systems couple voting power, physical node and computation power. This is no longer the case with stake: different nodes can have arbitrarily different stakes. This problem is compounded by the fact that stake is unbounded and often in the billions~\cite{algorand}. A single physical node can \revision{effectively}{thus, in effect,} carry both arbitrarily large or arbitrarily small stake. 

\revision{We}{In a nutshell, we} want to ensure that we maintain the same correctness and performance guarantees as in non-staked systems. Unfortunately, the round-robin approach we described in \S\ref{ss:failurefree} 
no longer works well. 
Consider for instance a system with $\n{i}=1000$ total stake, spread over two machines. $\Replica{i}{1}$ is Byzantine and has $\share{1}=\uf{i} = 333$, while $\Replica{i}{2}$ has $\share{2} = 667$. Using round-robin across these replicas disproportionately favors $\Replica{i}{1}$ which represents only $33.3\%$ of the shares in the system, yet is tasked with sending/receiving half the total messages. We must thus skew choosing sender-receiver pairs towards nodes with higher stake.  To highlight the challenges involved, we first sketch two strawmen designs:
\begin{itemize}[leftmargin=*, nosep, wide]
\item \textit{Version 1: Skewed Round-Robin.}  The most straightforward approach is to have \revision{replica $\Replica{il}{}$}{$l$-th replica} with stake \revision{$\share{il}$}{$\share{l}$} use round-robin scheduling to  send
\revision{$\share{il}$}{$\share{l}$} messages on its turn. This is, eventually, completely fair since all nodes send precisely as many messages as they have stake in the system. Unfortunately, this solution suffers from very poor performance under failure as it has \textit{no parallelism}: if stake is in the order of billions in the system, a single \revision{faulty}{malicious} node may\revision{}{, sequentially,} fail to send large contiguous portions of the message stream, triggering long message delivery delays. Rounding stake is unfortunately not an option: as stake is unbounded, each physical node can, in effect, have infinitely small (or arbitrarily large)  stake in the system. One physical node can have $\share{l}=1$ while another has $\share{l}=1\times10^9$. Rounding errors weaken liveness as more retransmissions may be needed to identify a correct sender-receiver pair.
\item \textit{Version 2: Lottery Scheduling.} For our next attempt, we consider lottery scheduling, a probabilistic scheduling algorithm~\cite{waldspurger}. Each node is allocated a number of tickets according to its stake; the scheduler then draws two random tickets to choose the next sender and the next receiver. Lottery scheduling addresses the parallelism concern mentioned above. Over long periods of time, the protocol is completely fair, and each sender-receiver sends/receives according to its stake. Unfortunately, due to the randomized nature of the protocol, over short periods of time, the proportion of sender and receiver pairs chosen may skew significantly from their shares in the system.
\end{itemize}

\par \textbf{Dynamic Sharewise Scheduler.} Our solution must (1) offer good parallelism; trustworthy replicas should be able to send messages in a bounded unit of time, (2) ensure fairness over both short and long periods; each node should send messages proportional to its shares, and (3) tolerate arbitrary stake values. These properties are exactly those that the Linux Completely Fair Scheduler (CFS) seeks to enforce. CFS defines a configurable time quantum during which each process is guaranteed to be scheduled; each process then gets CPU time proportional to its priority. 

Our \textit{dynamic sharewise scheduler} (DSS) adopts a similar strategy with one key modification. As stake is unbounded, DSS cannot guarantee as easily as CFS that all nodes will send a message within a fixed time period $t$. Instead, DSS maximizes the following objective: given a fixed time period $t$, how can \Scrooge{} schedule sender-receiver pairs such that each node sends/receive messages \textit{proportionally} to its shares. While this may appear straightforward, the ability for nodes to have arbitrarily large (or small) stake makes reasoning about proportionality challenging.
DSS turns to the mathematics of {\em apportionment} to handle this issue~\cite{apportionment,apportionment-math}.  Note that \Scrooge{} uses DSS to identify both senders and receivers in the same way. For simplicity, we thus discuss apportionment from the perspective of senders only. 

Apportionment is used to fairly divide a finite resource between parties with different entitlements or weights\footnote{It is, for instance, used to assign the number of seats per state in the US House of Representatives.}. Formally, an apportionment method $M$ defines a multivalued function $M(\vec{t},q)$. \revision{Here,}{} $\vec{t}$ represents the entitlement of node $\Replica{i}{l}$, that is the amount of messages that it should send or receive. In our case, this corresponds to its stake $\vec{t}_{il}=\share{il}$. 
$q$ denotes the total number of messages that can be sent in the specified time quantum $t$.  DSS makes use of Hamilton's method of apportionment~\cite{apportionment,apportionment-math}, which proceeds in four steps: 
\begin{itemize}[nosep,wide]
    \item First, DSS finds the standard divisor ($SD_i$), the ratio of the total amount of stake over the number of messages in a quantum, $SD_i = \frac{\pmb{\Delta_{i}}}{q}$. Intuitively, this defines how much stake must "back" each message. 
   \item Next, DSS computes the standard quota ($SQ_{il}$) for each node $\Replica{i}{l}$, $SQ_{il} = \frac{\share{il}}{SD_i}$, which indicates how many messages each replica should send. As this number may not be a whole number, DSS also computes the matching \textit{lower quota} ($LQ_{il}$), which takes the floor of the standard quota. The difference between the standard quota and the lower quota is \revision{called}{} the penalty ratio $PR_{il}$.
   \item DSS adds up these lower quotas to find the number of messages that will be sent $q_{whole} = \sum_l^{\n{i}}{LQ_{il}}$, without worrying about any unfairness introduced by rounding.
    \item If $q_{whole} < q$, that is if there is still space to send additional messages, DSS decides to increment the allocation of each $R_{il}$, in decreasing order of penalty ratio $PR_{il}$.
\end{itemize}

\par \textbf{Worked Example.} Intuitively, the algorithm described above splits messages fairly across nodes while minimizing the degree of imbalance introduced by the need to round stake up or down. Consider for instance the stake distribution and message quanta in Figure~\ref{table:apportionment}. The first two scenarios are straightforward as each replica has equal \revision{amounts}{amount} of stake. In both settings, running Hamilton methods, with a SD of $1$ in $d_1$ and of $10$ in $d_2$ reveals that each node should send $25$ messages.  $d_3$ highlights where apportionment shines. In this example, \revision{stake is}{stakes are} not distributed equally \revision{among}{amongst} replicas. The SD is $10$ as before. Replicas obtain $LQ$s respectively of $21$ for $\Replica{i}{0}$ 
($PR_{i0}=0.4$) and $26$ for the other three replicas $(PR_{i1}=PR_{i2}=PR_{i3}=0.2)$. The sum of all $LQ$ yields only $99$. As such, there is one message left to assign after considering the ``easily partitionable'' work. $\Replica{i}{0}$ has the highest $PR$ and is thus furthest away from a fair assignment. Hence, we increase its message assignment by $1$, from $21$ to $22$. 

\begin{figure}
    \scriptsize
    \centering
    \setlength{\tabcolsep}{5pt}
    \begin{tabular}{|c|c|c||c|c|c|c||c|c|c|c|}
    \hline
     DSS & Stake & q & $\share{0}$ & $\share{1}$ & $\share{2}$ & $\share{3}$ & $c_0$ & $c_1$ & $c_2$ & $c_3$  \\\hline
     $d_1$ & 100   & 100 & 25 & 25 & 25 & 25 & 25 & 25 & 25 & 25 \\ \hline
     $d_2$ & 1000 & 100 & 250 & 250 & 250 & 250 & 25 & 25 & 25 & 25 \\ \hline
     $d_3$ & 1000 & 100 & 214 & 262 & 262 & 262 & 22 & 26 & 26 & 26 \\ \hline
     $d_4$ & 100 & 10 &  97 & 1 & 1 & 1 & 10 & 0 & 0 & 0 \\ \hline
    \end{tabular}
\caption{Apportionment Example. $c_0,...c_3$ refers to the number of messages that must be sent (or received) by each node per quanta}
\label{table:apportionment}
\end{figure}

%
\subsection{Retransmissions}
\raf{Can someone look over this paragraph for writing quality? I wrote it and it feels kinda clunky. Also I don't think the idea that the "effective stake pairing when Replica $\Replica{il}{}$ sends a message to Replica$\Replica{jw}{}$ is $min(\share{il},\share{jw})$ which is the source of the excess resends}

Two issues remain to ensure eventual delivery with stake:
(1) the process of apportionment may select so few senders and receivers ($q<\uf{s}+\uf{r}+1$) that reliable delivery is not guaranteed.
(2) if the total stake across both \RSM{}s is large, then all safe $q > \uf{s}+\uf{r}+1$ may be too large to achieve parallelism. For example, if the total stake of \RSM{} $\SMR{s}$ is $\pmb{\Delta}_s = 4$ and \RSM{} $\SMR{r}$ is $\pmb{\Delta}_r = 4,000,000$ then 
$q>\uf{s}+\uf{r}+1 = 1,333,335$ which is an unrealistic number of messages to generate in a time quantum.

 The core issue present is that for reliable delivery, every message $m_\Seqn$, across all resends, must be sent and received by nodes whose stake, together, exceeds $\uf{s}+\uf{r}+1$. This  couples the number of resends needed to the (effectively unbounded) amount of stake in a network, and forces us to use increasingly large time quanta.  Consider two networks with identical large stake values. If $\SMR{s}$ and $\SMR{r}$ both have $\pmb{\Delta}_s=\pmb{\Delta}_r=4,000,000$, with each node having $1,000,000$ stake, each message send would pair replicas with $1,000,000$ stake and we would reach $\uf{s}+\uf{r}+1=2,666,667$ \revision{stake}{} after 3 message sends even without apportionment. This contrasts with our original example ($\pmb{\Delta}_s = 4$,$\pmb{\Delta}_r = 4,000,000$). Each replica in $\SMR{s}$ and $\SMR{r}$ is equally trusted, but we require $\uf{s}+\uf{r}+1 = 1,333,335$ resends solely because the \textit{relative} value of stake in the two $\RSM{}$s has changed.

Thankfully, this is not fundamental. To sidestep this issue, \Scrooge{} proportionally scales up the weights of the two communicating \RSM{s} to
their Least Common Multiple ($LCM$), and handles failures with the scaled stake values independent of apportionment.
For instance, assume that the total stake of \RSM{} $\SMR{s}$ is $\pmb{\Delta}_s$, \RSM{} $\SMR{r}$ is $\pmb{\Delta}_r$ and the $LCM = \lcm(\pmb{\Delta}_s, \pmb{\Delta}_r)$. 
\Scrooge{} scales the two \RSM{s} as follows:
\begin{enumerate}[nosep]
    \item Compute the multiplicative factor $\psi$ for each \RSM{}: $\psi_s = \frac{LCM}{\pmb{\Delta}_s}$ and $\psi_r = \frac{LCM}{\pmb{\Delta}_r}$.
    \item Multiply the stake of each replica with the multiplicative factor of its \RSM{}.
\end{enumerate}

Scaling up \RSM{s} is only necessary during message failures, allowing to keep message quanta small in the common-case. A replica thus uses the scaled up RSM weights upon receiving its first duplicate quack for a message $m$.

%% file: eval-micah.tex
\begin{figure}[t]
    \centering
     \begin{subfigure}[b]{0.32\columnwidth}
         \centering
         \includegraphics[width=0.935\textwidth]{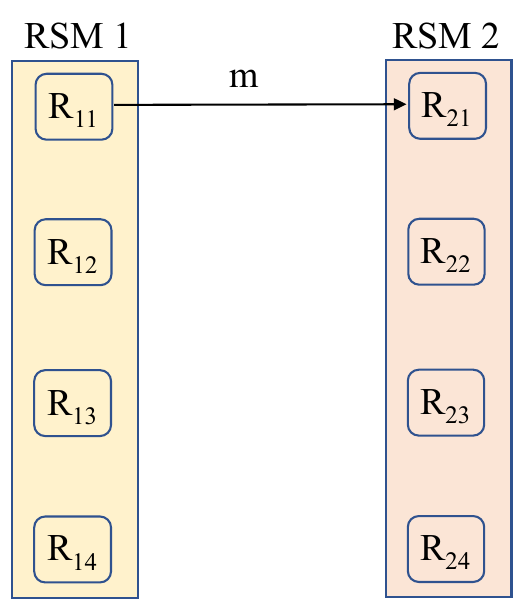}
         \caption{\OTO{}}
         \label{fig:one-to-one}
     \end{subfigure}
     \begin{subfigure}[b]{0.32\columnwidth}
         \centering
         \includegraphics[width=\textwidth]{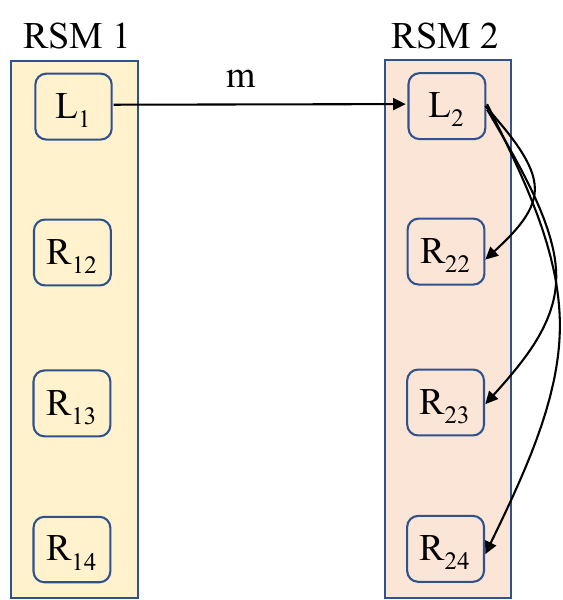}
         \caption{\LL{}}
         \label{fig:leader-to-leader}
     \end{subfigure}
     \begin{subfigure}[b]{0.32\columnwidth}
         \centering
         \includegraphics[width=0.935\textwidth]{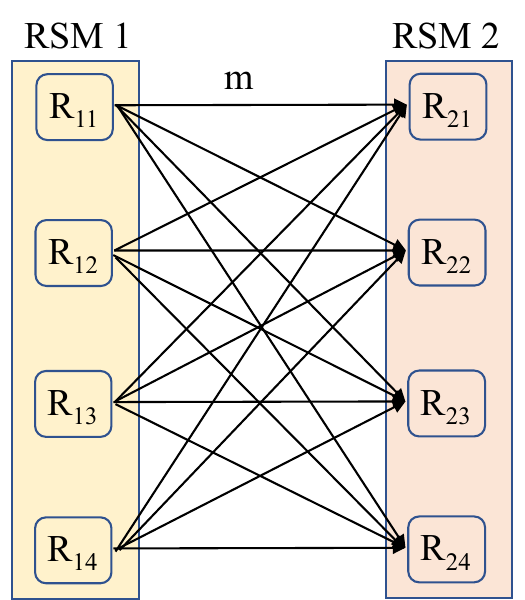}
         \caption{\ATA{}}
         \label{fig:all-to-all}
     \end{subfigure}
     \begin{subfigure}[b]{0.32\columnwidth}
         \centering
         \includegraphics[width=\textwidth]{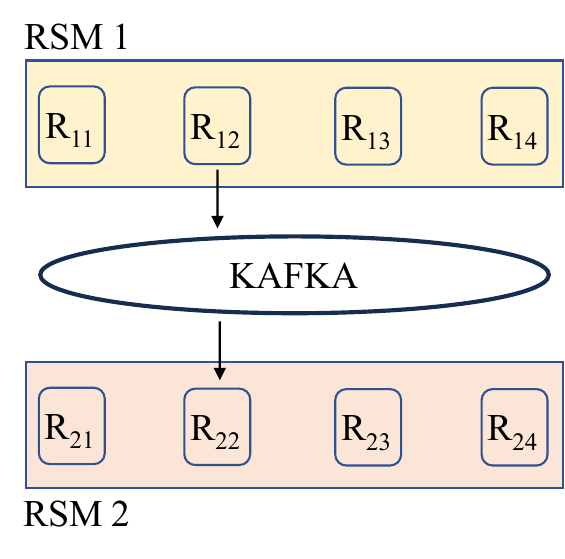}
         \caption{\Kafka{}}
         \label{fig:kafka}
     \end{subfigure}
     \begin{subfigure}[b]{0.32\columnwidth}
         \centering
         \includegraphics[width=\textwidth]{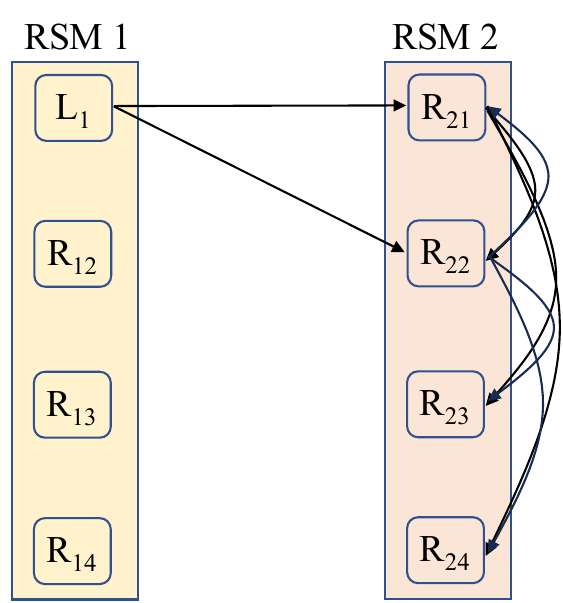}
         \caption{\OTF{}}
         \label{fig:geobft}
     \end{subfigure}
     \begin{subfigure}[b]{0.32\columnwidth}
         \centering
         \includegraphics[width=\textwidth]{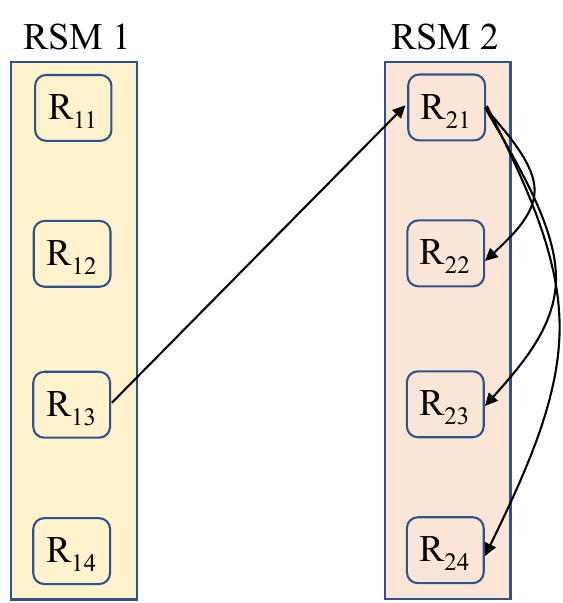}
         \caption{\Scrooge{}}
         \label{fig:scrooge}
     \end{subfigure}
    \caption{\revision{\CCC{} baseline summary.}{}}
    \label{fig:c3b-protocols}
\end{figure}

\begin{figure*}[!th]
    \centering
    \setlength{\tabcolsep}{1pt}
    \scalebox{0.6}{\ref{mainlegendN}}\\[5pt]
    \begin{tabular}{cccc}
    \graphFMsgB & \graphFMsgMB  & \graphFFTP  & \graphFSTP    
    \end{tabular}
    \caption{Throughput of \CCC{} protocols as a function of network size and message size}
    \label{fig:file-non-fail}
    \vspace{-6mm}
\end{figure*}

\section{Evaluation}
\label{s:eval}
\Scrooge{} aims to offer good performance in the common-case, while remaining robust to faults when failures do arise. We aim to answer the following three questions.
\begin{enumerate}[nosep,wide]
    \item How does \Scrooge{} perform in the common case (\S\ref{ss:file})?
    \item How does \Scrooge{} remain robust to failures (\S\ref{ss:eval:failures})?
    \item How does \Scrooge{} perform in real applications (\S\ref{ss:real-world})?
\end{enumerate}


\par \textbf{Implementation} We implemented \Scrooge{} in $\approx$ 4500 lines of C++20 code with Google Protobuf v$3.10.0$ for serialization and NNG v$1.5.2$ for networking~\cite{picsou-artifact}.
\Scrooge{} is designed to be a plug-and-play library that can be easily integrated with existing \RSM{s}.
 We evaluate \Scrooge{} against five other comparable protocols (Figure~\ref{fig:c3b-protocols}).
\begin{enumerate}[nosep,wide]
    \item {\bf One-Shot (\OTO{})}: In OST, a message is sent by a single sender to a single receiver. 
    \OTO{} is \textit{only} meant as a performance upper-bound. It does not satisfy \CCC{} as message delivery cannot be guaranteed.   
    \item {\bf All-To-All (\ATA{})}: In \ATA{}, every replica in the sending \RSM{} sends all messages to all receiving replicas ($O(\n{s}\times \n{r})$ message complexity). Every correct receiver is guaranteed to eventually receive the message.
    \item {\bf Leader-To-Leader (\LL{})}: The leader of the sending RSM sends a message to the leader of the receiving RSM, who then internally broadcasts the message. 
    This protocol does not guarantee eventual delivery when leaders are faulty. 
    \item {\bf \Kafka{}}: 
    Apache \Kafka{} is the de-facto industry-standard for exchanging data between services~\cite{kreps2011kafka}. Producers write data to a Kafka cluster, while consumers read data from it. Kafka, internally, uses Raft~\cite{raft} to reliably disseminate messages to consumers. We use Kafka 2.13-3.7.0.
    \item {\bf \OTF{}}: GeoBFT \cite{geobft,blockchain-book}  breaks down an RSM into a set of sub-RSMs. Much like \LL, GeoBFT's cross-RSM communication protocol, \OTF{}, has the leader of the sender \RSM{} send  its messages to at least $\uf{r}+1$ receiver \RSM{} replicas. Each receiver then internally broadcast these messages. When the leader is faulty, replicas timeout and request a resend. \OTF{} thus guarantees eventual delivery after at most $\uf{s}+1$ resends in the worst-case (for  $O(\uf{r} * \uf{s})$ total messages).
\end{enumerate}

\par \textbf{\RSM{s}.} We consider four representative RSMs.
\begin{enumerate}[nosep,wide]
    \item \textbf{\File{}}: An in-memory file from which a replica can generate committed messages infinitely fast. This is a baseline to artificially saturate the \CCC{} protocols.
    \item \textbf{\Raft{}~\cite{etcd-raft}}: A widely used CFT \RSM{}, used in services like Kubernetes Cluster. We run Etcd's \Raft{} version v3.0. 
    \item \textbf{\ResDB{}~\cite{resilientdb}}: A high performance implementation of \textit{PBFT}~\cite{pbftj}, a well-known representative BFT protocol.
    \item \textbf{\Algo{}~\cite{algorand}}: A  popular \PoS{} blockchain protocol~\cite{algorand}. 
\end{enumerate}

\par {\bf Experimental Setup.}
We deploy up to $45$ GCP c2-standard-8 nodes 
(Intel Cascade Lake, $8$vCPU, $32$ GiB RAM, 15 GBits/s). 
Each experiment runs for $180$ seconds ($30$ second warmup/cool down). All experiments run \Scrooge{} with a $\phi$-list of \si{200}{k} and $256$ bits for 0.1kB and \si{1}{MB} \revision{messages}{batches}, respectively (best results for our specific network setup).  We further assume that RSMs forward all messages to the other RSM, as this represents a worst-case scenario for \Scrooge{}. As is standard~\cite{hotstuff,quepaxa,basil,kauri,rcc}, \revision{unless stated otherwise}{}, replicated operations in our experiments are no-ops \revision{}{only}, which ensures that the bottleneck is \revision{not}{never} execution.

\par {\bf Metrics.}
\textit{\RSM{} throughput} is the number of consensus invocations completed at an \RSM{} per second; 
\textit{\CCC{} throughput} is the number of completed \CCC{} invocations per second.
When baselines, like \OTO{}, do not acknowledge received messages,
we calculate \CCC{} throughput as the number of unique messages sent from sender \RSM{} to receiver \RSM{}.

\subsection{File \RSM{} Common Case Performance}
\label{ss:file}



Our first set of experiments aim to stress test the six \CCC{} protocols (\Scrooge{}, \OTO{}, \ATA{}, \LL{}, \OTF{}, and Kafka) \textit{without} failures. We use the ``infinitely fast'' File RSM to saturate all \CCC{} implementations. In all cases, we include the \OTO{} line as the upper-bound of our networking implementation.

{\bf Varying number of replicas in each \RSM{}.}
We first consider the relative performance of \Scrooge{} as a function of the network size.
We fix the message size to \SI{0.1}{kB} and \SI{1}{MB} and increase the number of replicas in each \RSM{} from $4$ to $19$ (Figure~\ref{fig:file-non-fail}~(i)-(ii)). For small network sizes, \Scrooge{} outperforms \ATA{} by a factor of $2.5\times$ (small messages) and $3.2\times$ (large messages) and in larger networks it increases to $6.6\times$ and $12.1\times$. \Scrooge{} sends only a linear number of messages, while \ATA{} must send a quadratic number of messages.
Like \Scrooge{}, \LL{} and \OTF{} send a linear number of messages, but quickly bottleneck at the leader since it needs to send every message.
\OTO{}'s performance, as expected, increases with network size as increasing the number of replicas increases the number of parallel messages.
Kafka performs significantly worse in all cases, as it internally runs consensus. 
\par{\bf Varying Message Size.}
In Figure~\ref{fig:file-non-fail}~(iii)-(iv), we fix the size of each \RSM{} to $n=4$ (small) and $n=19$ replicas (large)
and increase the message size from \SI{0.1}{kB} to \SI{1}{MB}.  As expected, the performance of each \CCC{} implementation drops as a linear function of the message size. Note that \Scrooge{} performs relatively better than other protocols for large message sizes  as they hide the moderate compute overheads introduced by the system. 
For instance, on a large network \Scrooge{} performs over $12\times$ better than \ATA{}, \LL{}, and \OTF{} for large messages. Instead, for small messages, \Scrooge{} only performs $6.6\times$, $4.4\times$, and $4.9\times$ better (respectively).  

\par \textbf{Impact of Stake} Next, in Figure~\ref{fig:file-geo-stake}~(i), we study how well \Scrooge{} performs for weighted \RSM{s} when stake distribution becomes unequal. We fix the message size to \SI{100}{B}.


\raf{This seems to be repetitive with the next paragraph...} Consider 1)~two \RSM{s} where the throughput is \textit{throttled} and one replica in each \RSM{} gets increasingly more stake;
2)~two \RSM{s} where throughput is not throttled, but one replica still gets a larger share of stake over time. 
Our aim is to demonstrate that \Scrooge{} does not lose any performance under unequal stake distributions. 

We run two experiments. First, we artificially throttle the \revision{File \RSM{}}{available bandwidth} such that \Scrooge{} \revision{cannot transmit over 1M txns/s}{transmits 1M txn/s}, regardless of the stake distribution (flat 1M lines on the graph). Next, we allow each node to have access to \revision{an unmodified File \RSM{}}{all the available bandwidth}. In these experiments \Scrooge~i \raf{\Scrooge~i also does the same in the first experiment which currently is unclear} refers to the setting where the \revision{high-stake}{bottlenecked} node has $i \times$ more stake than other nodes. Initially, shifting the stake distribution to one node does not affect performance as the high stake node can handle the additional load. Eventually, however, this node becomes a bottleneck, thus causing throughput to decrease.

\par \textbf{Geo-replication}  \revision{In}{Finally, in} Figure~\ref{fig:file-geo-stake}(ii), we run geo-replicated experiments by deploying one \RSM{} in US-West and \revision{the}{} other \RSM{} in Hong Kong
(cross-region bandwidth, pair-wise is \SI{170}{Mbits\per\sec}, \revision{RTT}{ping time} \SI{133}{ms}). We fix the message size to \SI{1}{MB} and vary \RSM{} size from $4$ to $19$. The lower bandwidth across \revision{pairs of machines}{\RSM{}s} disproportionally affects \ATA{}, \LL{}, and \OTF{}.
\Scrooge{} outperforms \ATA{}, 
by $12\times$ 
(for network size $4$) and $44\times$, 
(for network size $19$). Somewhat counter-intuitively, the performance of both \Scrooge{} and \OTO{} increase as a function of network size; increasing the number of receivers gives senders access to more bandwidth in Google Cloud. \Scrooge{} intentionally has its senders send to multiple receivers and thus (artificially) outperforms \OTO{}, which fixes unique sender-receiver pairs.

\begin{figure}
    \centering
    \setlength{\tabcolsep}{1pt}
    \scalebox{0.6}{\ref{mainlegendGEO}}
    \scalebox{0.6}{\ref{mainlegend2}}\\[5pt]
    \begin{tabular}{cc}
    \graphFStake & \graphGEO 
    \end{tabular}
    \caption{Impact of Stake and Geo-replication.}
    \label{fig:file-geo-stake}
\end{figure}

\subsection{Impact of failures}
\label{ss:eval:failures}
We now consider performance under failures.

\begin{figure*}
    \centering
    \setlength{\tabcolsep}{1pt}
    \scalebox{0.57}{\ref{mainlegendCrash}} %
    \scalebox{0.57}{\ref{mainlegend5}} 
    \scalebox{0.57}{\ref{mainlegend4}} \\[5pt]
    \begin{tabular}{ccc}
    \graphFail \quad   & \graphBPhiList \quad  &   \graphBAcking 
    \end{tabular}
    \caption{Effects of Failures on \Scrooge{}.}
    \label{fig:byz-failures}
    \vspace{-6mm}
\end{figure*}
\par \textbf{Crash Failures.} 
In this experiment,  we \revision{}{randomly} crash 33\% of the replicas in each RSM (Figure~\ref{fig:byz-failures}~(i)); message size set to \SI{1}{MB} \revision{}{(similar results for smaller messages)} and $\phi$-list size as \revision{256}{$512$}.
\Scrooge{}'s performance drops by a factor of $22.8\%-30.5\%$. This is expected:
\Scrooge{}, by default, fully maxes out links with "useful" information. Removing a third of the links thus removes a third of the available bandwidth. 
Nonetheless, \Scrooge{} continues to outperform \ATA{}, \OTF{}, and \LL{} by at least $2\times$ on small networks, and up to $8.9\times$ on larger networks.


\par \textbf{Byzantine Failures.}
Next, we consider the impact of Byzantine failures in the system.
While it is impossible to model all arbitrary failures, we consider four main classes of attacks. Malicious nodes can (1) send invalid, uncommitted messages, 
(2) collude to drop long sequences of messages $\uf{s}+\uf{r}$ times, 
(3) selectively drop messages, and (4) send incorrect acknowledgments. The first attack amounts to a DDOS attack (as correct replicas will discard invalid messages) and is thus out of scope. \Scrooge{} defends against the second attack by assigning node IDs using a verified source of randomness
(the probability that all byzantine nodes get assigned contiguous node IDs is negligible).
We focus on the last two scenarios.

\begin{enumerate}[nosep,wide]
    \item \textbf{Impact of $\phi$-list scaling on Byzantine failures.}
    $\phi$-lists bound the possible performance drop from malicious nodes selectively dropping messages.
    We again assume $33\%$ \revision{of replicas are faulty}{faulty replicas} in both RSMs (Figure~\ref{fig:byz-failures}~(ii)), this time Byzantine. We consider a message size of \SI{1}{MB}. 
    Our results illustrate that the larger $\phi$-list size helps \Scrooge{} quickly recover from Byzantine failures, despite the larger $\phi$-list increasing \revision{metadata}{message} sizes.
    We observe that a $\phi$-list size of $256$ is optimal for recovering from the $33\%$ Byzantine attacks. As the network gets larger, the time it takes to complete a full broadcast gets longer, which increases \revision{the}{} latency to confirm a delivery. Thus, more messages can be dropped before we can detect that they are dropped, hence the larger $\phi$-list.
    \item \textbf{Sending incorrect acks.} 
    Malicious nodes can choose to lie in their acknowledgments. We simulate this behavior in Figure~\ref{fig:byz-failures}~(iii) by having malicious nodes send acks for overly high sequence numbers (Picsou-Inf), overly low ones (Picsou-0) or offset by $\phi$ (Picsou-Delay).
    We find that this behavior is much less harmful than simply crashing.
    Correct nodes wait for a quorum of $\uf{r}+1$ matching acks in order to consider the message delivered, and thus already assume that $\uf{}$ of those acks will be lies. 
    Lying about an ack thus only temporarily delays the formation of a quorum.  
\end{enumerate}

\subsection{Application Case Study}
\label{ss:real-world}
We now study impact on real-world applications (Section~\ref{s:intro}).

\textbf{Disaster Recovery.} 
Disaster recovery (DR) ensures continued fault-tolerance in the presence of full datacenter outages, and is a popular feature of modern cloud environments~\cite{google-dr,aws-dr,azure-dr,disaster-recovery-confluent}. 
DR deployments often implement cross-datacenter \RSM{} mirroring over Kafka, where the Kafka cluster is located in the receiving datacenter. 
We run Etcd DR~\cite{disaster-recovery-confluent} by deploying two Etcd \RSM{s} in two distinct datacenters, one in \revision{GCP}{} region us-west-4 and \revision{the}{} other in us-east-5. 
Communication is unidirectional for DR, since only a single sending \RSM{} is sending data to the mirrored \RSM{} and the mirrored \RSM{} does not have any information to send back (other than acks).

Etcd DR invokes \Scrooge{} on all {\em put} transactions and assigns them a new, sequential, internal sequence number. This new sequence number is necessary as DR only applies to a subset of \revision{Etcd}{} transactions (just puts, not gets \revision{or reconfiguration}{}). \revision{}{\Scrooge{} requires contiguous sequence numbers in order to count \quack{s} properly.} 
The receiving \RSM{} thus simply applies all put transactions in sequence number order.

In Figure~\ref{fig:case-study}~(i), we plot the throughput of Etcd DR \revision{(in MB/s)}{} with various \CCC{} protocols 
for different message sizes; each \RSM{} has $5$ replicas.
\OTO{} achieves maximum theoretical throughput \revision{for an Etcd cluster running}{ from} a \CCC{} protocol; \ETCD{} 
is the baseline for maximum throughput from a single Etcd \RSM{} without any communication; one can only transmit messages as fast as Etcd commits them. 
There are two primary resource bottlenecks in the system: the cross-region network bandwidth and Etcd's disk goodput (since it synchronously writes each transaction it commits to disk). \ATA{} broadcasts every message to all machines, so its throughput is bottlenecked by the cross-region network bandwidth (50 MB/s). 
Similarly, \OTF{} and \LL{} are bottlenecked because they limit the number of nodes sending unique messages over the network in parallel. 
In contrast, \Scrooge{} shards the set of messages across all sending nodes, so each node uses 50 MB/s bandwidth to send $1/5$-th of the messages ($5$ nodes per \RSM{}). Thus, \Scrooge{} has an effective 250 MB/s of bandwidth available, resulting in saturating Raft's disk goodput of 70 MB/s. 
In case of \Kafka{}, we can only deploy $3$ nodes, at most $3$ shards, so it can achieve at most 150 MB/s.
\Kafka{} can still can achieve potentially the same goodput as \Scrooge{}. 
However, in our testing, \Kafka{} was still unable to achieve optimal performance given its sensitivity to high network latency.

\textbf{Data Sharing and Reconciliation}
As described in \S\ref{s:intro} (Figure~\ref{fig:case-study}(ii)), there are operational and sovereignty concerns associated with managing a single RSM across trust domains. We implement the data reconciliation application described in~\cite{ccf}. In this setup, two distinct entities, Agency A and Agency B, run their own Etcd \RSM{} but exchange data to ensure that any shared state remains consistent. Specifically, each \RSM{} sends {\em key-value updates} for shared data. The receiver then checks whether the values match and takes remedial action if not. Communication between RSMs is bidirectional\revision{}{in this case}.
\ATA{}, \LL{}, \OTF{}, and \Scrooge{} all behave similarly to the performance discussed in the disaster recovery experiment, albeit with a lower starting goodput since there is extra processing time needed for \revision{looking up keys and comparing their values}{the receiving \RSM{} in the CCF application}. \Kafka{} had unusually low performance since we were running into a known issue with high latency \Kafka{} consumers which are not addressed in these result. We are in the process of addressing this.

{\bf Decentralized Finance.} Our final application implements a blockchain bridge, designed to foster interoperability between chains~\cite{ccip}, for instance for asset transfer. We implement an asset transfer application across three types of wallets: (1) two \PoS{} \Algo{} chains, (2) two traditional permissioned PBFT \ResDB~\cite{resilientdb,geobft} chains, and 
(3) interoperability between \ResDB{} to \Algo{} chains.
\Algo's base throughput with another \Algo~instance is 120 blocks/second. \ResDB's base throughput when communicating with another \ResDB~cluster is $\approx$6000 batches/second (of size \si{5}{kB}). The cross-chain throughput when \Algo~sends to \ResDB~is 135 blocks/second.
\Scrooge{} has minimal impact on the throughput of any of the \RSM{s}, with less than 15\% decrease in throughput in the worst case. \revision{This decrease in throughput is independent of node stake.
Latency will instead increase proportionally to network size -- this property is fundamental to \Scrooge{}'s high throughput, but may be unacceptable in some large scale blockchain or \RSM{} deployments.}{}
(2) \Scrooge{} successfully handles throughput differences between \RSM{s}; 
the slow \Algo{} \RSM{} efficiently communicates with the much faster \ResDB{} \RSM{}.



%% file: related.tex
\section{Related Work}
\begin{figure}[!t]
    \centering
    \setlength{\tabcolsep}{1pt}
    \scalebox{0.6}{\ref{mainlegendApps}}\\[5pt]
    \begin{tabular}{cc}
    \hspace{-8mm}
    \graphDR & \graphCCF
    \end{tabular}
    \caption{Disaster Recovery and Data Reconciliation.}
    \label{fig:case-study}
\end{figure}
The problem of reliably sending messages \revision{within}{between} groups of participants through \textit{reliable broadcast} or \textit{group communication}
is  well-studied~\cite{fault-tolerant-broadcast,reliable-birman,bracha-toueg-ba,sintra,diffusion-byzantine-env,broadcast-survey,hadzilacos-thesis,delta-reliable-broadcast}, in both the CFT and BFT setting ~\cite{random-ba,bracha-ba,bracha-toueg-ba,sintra,good-byzantine-broadcast,synchronous-ba-ittai}. These works consider communication among groups but do not consider communication between groups. \Scrooge{} leverages the internal guarantees provided by these communication primitives to build a group-to-group communication primitive, C3B.

\par \textbf{Logging Systems.} Shared logs are a popular way for reliably exchanging messages~\cite{kreps2011kafka,kalia2016design,jia2021boki,cao2018polarfs,balakrishnan2013tango,wang2015building,venkataraman2017drizzle,fuzzylog,scalog}. Systems such as Kafka~\cite{kreps2011kafka}, RedPanda~\cite{redpanda}, Delos~\cite{delos} have become industry standards~\cite{delos}.
While these systems work well in the CFT setting, they are not directly applicable to the \BFT{} setting: this log becomes a central point of attack. Moreover, most of these systems use relatively heavyweight fault tolerance: Kafka, for instance, internally makes use of Raft.


\par \textbf{Communication between RSMs.} Two lines of work have considered communication between RSMs, but in different contexts. First, Aegean~\cite{aegean} makes a similar observation as this paper: it highlights that replicated services rarely operate in a vacuum and must instead frequently communicate. 
However, Aegan solves a strictly orthogonal problem. It focuses on how to correctly replicate services that can issue nested requests to other (possibly replicated) services. Aegean presents the design of a shim layer that exists between replicated service and backend service and manages all the communication/data storage. Second, Byzantine fault tolerant communication between \RSM{s} has been a topic of interest in the context of \textit{sharded} BFT systems that view each shard as an independent \RSM{}. These shards periodically need to communicate with each other to process cross-shard transactions~\cite{ahl,sharper,ringbft,rapidchain,basil,conflux,blockvsdist,txallo,blockchain-book}.
Most of these systems simply adopt the all-to-all communication pattern between the shards that we evaluate in \S\ref{s:eval}. 
\revision{GeoBFT~\cite{geobft} and Steward~\cite{steward} are two exceptions.
Steward uses a hierarchical consensus architecture; all communication between the clusters is managed by a designated primary cluster, which internally replicates requests via Paxos. GeoBFT uses \OTF{}.}{GeoBFT~\cite{geobft} is the exception; the leader sends each message to $\uf{}+1$ replicas in the receiving \RSM{} (total messages sent is thus O($\n{}$)). Similarly to \Scrooge{}, each receiver broadcasts each message received from the other \RSM{} in its \RSM{}. This is still $O(\n{})$ times more messages than \Scrooge's (w.h.p) $O(1)$ strategy.}


\par \textbf{Blockchain bridges.} With the rise of blockchain technology and cryptocurrencies~\cite{splitbft,rbft,basalt,bitcoin-johnnatan,nakamoto-narula,epaxos-revisit,simulation-blockchain,maria-thesis,bc-processing,hotstuff-1}
there is a new found interest in blockchain interoperability~\cite{atomic-cross-chain-swap,blockchain-interop-survey,pow-sidechains,trustboost,sok-cross-chain,analyze-inter-blockchain-communication,blockchain-interoperability-survey, ccip}.  These works focus on the {\em correct} conversion of assets from one blockchain to the other. They can be broadly clustered into two groups (1) {\em blockchain bridges}, and (2) {\em trusted operators}. A blockchain bridge requires a replica of the sending \RSM{} to send a committed contract to a replica of the receiving \RSM{}. Recently, several such blockchain bridges have popped up~\cite{polynetwork,rainbrow-bridge,axelar-bridge}. Unfortunately, they provide few formal
guarantees, which has led to massive financial attacks and hacks~\cite{sok-cross-chain,zkbridge,trustboost}. Moreover, 
these bridges continue to be impractical because of their high cost~\cite{zkbridge}.
Trusted operator systems are, in contrast, much more practical
~\cite{polkadot,cosmos,blockchain-interoperability-survey}, but as the name suggests, they require centralized management. \revision{Works like Thema~\cite{Thema} instead use \BFT{} \RSM{}s to communicate between two non-replicated services.}{}

%% file: conclusion.tex
\section{Conclusion}

This paper introduces the \CCC{} primitive and proposes \Scrooge{}, an efficient implementation of \CCC{}. We show that, by borrowing techniques from TCP and adapting these to the crash and \BFT{} context, we can develop a solution that allows RSMs to efficiently exchange messages.

%% file: proofs.tex
\newpage

\section{Appendix}
\label{s:appendix}
Having described the \Scrooge{} protocol, we now prove its correctness.

\subsection{Correctness}
\label{ss:correctness}

\Scrooge{} guarantees safety and liveness under asynchrony. 
We first try to show that \Scrooge{} upholds the {\em eventual delivery} and {\em integrity} properties 
expected of a \CCC{} protocol.
\begin{lemma}\label{th:max-retransmit}
    {\bf Retransmission Bound.} At most $\uf{s}+\uf{r}+1$  retransmission attempts are required to successfully send a message $m$ from 
    \RSM{} $\SMR{s}$ to \RSM{} $\SMR{r}$, 
    given that at most $\rf{s}$ replicas of $\SMR{s}$ and $\rf{r}$ replicas of $\SMR{r}$ are Byzantine.
\end{lemma}
\begin{proof}
    In \Scrooge{}, for each message $m$, there is a predetermined initial sender-receiver pair and 
    subsequent sender-receiver pairs until a correct replica of \RSM{} $\SMR{r}$ delivers $m$.
    A failed send for $m$ is a result of the sender at $\SMR{s}$ and/or receiver at $\SMR{r}$ failing or acting Byzantine;
    replicas of \RSM{} $\SMR{s}$ detect a failed send for $m$ through duplicate \quack{s} (\S\ref{ss:failures}).
    \mk{I'm not following the logical flow here.}
    \nc{I'm not either}
    \nc{This proof only appears to apply to the non stake case.}

    To increase the number of retransmissions for message $m$, 
    an adversary needs to ensure that each faulty replica is paired to a correct replica; 
    each sender-receiver pair can force only one message retransmission irrespective of whether the sender, receiver, or both are faulty.
    Thus, following is a worst-case mapping for \Scrooge:  
    the first $\uf{s}$ senders of $m$ are Byzantine, are paired to correct replicas of $\SMR{r}$, and 
    decide to not send $m$ to these correct receivers.
    This results in $\uf{s}$ retransmissions for $m$.
    The following $\uf{r}$ senders of $m$ are correct, but are paired with Byzantine replicas of $\SMR{r}$, which decide to 
    ignore messages from $\SMR{r}$. 
    This results in an additional $\uf{r}$ retransmission for $m$.
    As we have iterated over all the Byzantine replicas in both the \RSM{s}, the $(\uf{s}+\uf{r}+1)$-th retransmission of $m$ is 
    guaranteed to be between a correct sender and a correct receiver, which
    ensures that $m$ is eventually delivered.
\end{proof}

\begin{theorem} \label{th:liveness}
    {\bf Liveness.} \Scrooge{} satisfies eventual delivery:
    If \RSM{} $\SMR{s}$ transmits message $m$ to \RSM{} $\SMR{r}$, then $\SMR{r}$ will eventually deliver $m$.
\end{theorem}
\begin{proof}
    For \Scrooge{} to satisfy eventual delivery: 
    it is sufficient to show that the following two cases hold:
    (1) if a correct replica in \RSM{} $\SMR{s}$ sends a message $m$, then eventually a correct replica in \RSM{} $\SMR{r}$ receives $m$.
    (2) if a sender for message $m$ is faulty, then eventually $m$ will be sent by a correct sender. 

    {\bf Case 1: Correct Sender.}
    
    If both the sender and receiver for message $m$ are correct, then \Scrooge{} trivially satisfies eventual delivery.
    This is the case because any message sent by a correct sender to a correct receiver, 
    will be received and broadcasted by the correct receiver. 
    This will ensure that $m$ is eventually delivered by $\SMR{r}$.
    
    Next, we consider the case when the receiver is faulty.
    We prove via induction on sequence number $\Seqn$ of message $m$ that if a correct replica of $\SMR{s}$ sends $m$, 
    then then eventually a correct replica in \RSM{} $\SMR{r}$ receives $m$. 
    
    {\em Base case ($\Seqn = 1$):} 
    We start by proving that the first message (sequence number $\Seqn = 1$) sent by a correct replica (say $\Replica{s}{i}$) 
    is eventually received by a correct replica in $\SMR{r}$.
    As $m$ is the first message to be sent, the last \quack{ed} message at each replica in $\SMR{s}$ is set to $\bot$.
    From Section~\ref{ss:failures}, we know that replicas in $\SMR{s}$ detect a failed send of a message through duplicate $\quack{s}$
    for the preceding \quack{ed} message (message with one less sequence number). 
    In the case of $m$, $\Seqn=1$, a replica will detect a failed send of $m$ from a duplicate \quack{} for $\bot$ and 
    identify the subsequent sender-receiver pair for $m, k=1$.
    From Lemma~\ref{th:max-retransmit}, we know that at most $\uf{s}+\uf{r}$ sender and receiver pairs can have one faulty node.
    This implies that $(\uf{s}+\uf{r}+1)$-th retransmission of $m, \Seqn=1$ will be between a correct sender-receiver pair.
    This correct receiver will broadcast $m, \Seqn=1$ to all the replicas of $\SMR{r}$.

    {\em Induction hypothesis:}
    We assume that replicas in $\SMR{r}$ have received all messages till sequence number $k-1$.

    {\em Induction case:}
    For induction step, we show that if a correct replica of $\SMR{s}$ sends $m, \Seqn$ (sequence number for $m$ is $\Seqn$), 
    then then eventually a correct replica in \RSM{} $\SMR{r}$ receives $m, \Seqn$.
    Replicas of $\SMR{s}$ will detect a failed send for $m, \Seqn$ when they receive a duplicate \quack{} for $m, \Seqn-1$. 
    As at most $\rf{s}+\rf{r}$ sender-receiver pairs can have a faulty node,
    $(\uf{s}+\uf{r}+1)$-th retransmission of $m, \Seqn$ will be between a correct sender-receiver pair, which 
    satisfies the validity property.

    {\bf Case 2: Faulty Sender.}
    
    Next, we consider the case when message $m$ is assigned to a faulty replicas, which did not communicate $m$ from $\SMR{s}$ to $\SMR{r}$.
    We prove this lemma through induction over the sequence number $\Seqn$ of message $m$.

    {\em Base case ($\Seqn = 1$):}
    We start by proving the base case for first message $m$ (sequence number $\Seqn = 1$).
    As $m$ is the first message to be communicated, the last \quack{ed} message at each replica in $\SMR{s}$ is set to $\bot$.
    Assume $m$ is assigned to a faulty replica, which did not communicate $m$ from $\SMR{s}$ to $\SMR{r}$.
    If this is the case, then eventually all the correct replicas of $\SMR{s}$ will receive a duplicate \quack{} for $\bot$ and 
    will detect a failed send for $m$.  
    As proved in Lemma~\ref{th:max-retransmit}, this will lead to another sender-receiver pair for $m, \Seqn=1$.
    If the sender is again faulty, then the correct replicas of $\SMR{s}$ will receive another duplicate \quack{} for $\bot$, 
    which will lead to yet another sender-receiver pair for $m, \Seqn=1$.
    This process will continue for at most $\uf{s}$ times as $(\uf{s}+1)$-th sender is guaranteed to be correct.
    
    {\em Induction hypothesis:}
    We assume that all the messages till sequence number $\Seqn-1$ have been eventually sent by correct replicas in $\SMR{s}$.
    Additionally, using Theorem~\ref{th:liveness}, we can conclude that all the messages till sequence number $\Seqn-1$ have been delivered by $\SMR{r}$.

    {\em Induction case:}
    Next, we show that a message $m$ with sequence number $\Seqn$ to be communicated from $\SMR{s}$ to $\SMR{r}$ 
    is eventually sent by a correct replica in $\SMR{s}$.
    As stated above, if the initial sender is correct, then the base case holds.
    Otherwise, $m, \Seqn$ is initially assigned to a faulty sender.
    Similar to the base case, eventually all the correct replicas of $\SMR{s}$ will receive a duplicate \quack{} for $m, \Seqn-1$ and 
    will detect a failed send. 
    Following this, $m, \Seqn$ will be eventually resent by a correct replica in at most $\uf{s}$ attempts.
\end{proof}


\begin{theorem}
    {\bf Safety.} \Scrooge{} satisfies the integrity property, that is, \RSM{} $\SMR{r}$ only delivers a message 
    only if it was transmitted by $\RSM{}$ $\SMR{s}$.
\end{theorem}
\begin{proof}
    We prove this theorem through contradiction by starting with the assumption that \RSM{} $\SMR{r}$ delivers a message $m$ that 
    was not transmitted by $\RSM{}$ $\SMR{s}$.
    If this is the case, then at least one of the replicas delivering $m$ was correct because \Scrooge{} marks a message delivered 
    if the replicas of the sending \RSM{} $\SMR{s}$ receive acknowledgment for $m$ from $\uf{r}+1$ replicas (at least one correct replica).
    A correct replica of $\SMR{r}$ will only acknowledge a message $m$ if $m$ carries a proof that proves $m$ was
    committed by a quorum of replicas in $\SMR{s}$.
    However, \RSM{} $\SMR{s}$ never transmits $m$, which means that $m$ was sent by a malicious replica in $\SMR{s}$ and 
    $m$ does not have support of a quorum of replicas.
    As a result, a correct replica will never acknowledge $m$, no sender will receive $\uf{r}+1$ acknowledgements, and 
    $m$ will not be marked delivered, which contradicts our assumption.
\end{proof}


\subsection{Bounds}
\label{subsec:bounds}

\par \textbf{Theorem 1.} \textit{
 Given a sending \RSM{} with $\n{s}= \alp{s} \times \uf{s} + 1$ replicas and 
receiving \RSM{} with $\n{r}= \alp{r} \times \uf{r} + 1$ replicas, where $\alp{s}, \alp{r} > 1$ are the replication factors of the \RSM{s},
\Scrooge{} needs to resend a message at most $72$ times to guarantee a $10^{-9}$ failure probability, irrespective of the number of nodes and failures.
}

\begin{proof}
The maximum number of faulty pairs (either the receiver or the sender is faulty) in this system of two \RSM{s} is:
\begin{equation}
Faulty = \uf{s} \times \n{r} + \uf{r} \times \n{s} - \uf{s} \times \uf{r}   
\end{equation}
By assumption, we have $\uf{s} = \frac{\n{s}-1}{\alp{s}}$ and $\uf{r} = \frac{\n{r}-1}{\alp{r}}$.
We get:
\begin{equation}\label{eq:2}
\begin{split}
Faulty  & = \frac{\n{s} \times \n{r}}{\alp{s}} + \frac{\n{r} \times \n{s}}{\alp{r}} - \frac{\n{s} \times \n{r}}{\alp{s} \times \alp{r}} \\
        & - \frac{\n{r}}{\alp{s}} - \frac{\n{s}}{\alp{r}} + \frac{\n{s} + \n{r}}{\alp{s} \times \alp{r}} - \frac{1}{\alp{s} \times \alp{r}}
\end{split}
\end{equation}
Given that $\alp{s}, \alp{r} > 1$, which is typical for any fault-tolerant \RSM{}, the following holds. 
\begin{equation}\label{eq:3}
    - \frac{\n{r}}{\alp{s}} - \frac{\n{s}}{\alp{r}} + \frac{\n{s} + \n{r}}{\alp{s} \times \alp{r}} - \frac{1}{\alp{s} \times \alp{r}} < 0
\end{equation}
From Equations~\ref{eq:2} and~\ref{eq:3}, we have:
\begin{equation}\label{eq:4}
Faulty  < \frac{\n{s} \times \n{r}}{\alp{s}} + \frac{\n{r} \times \n{s}}{\alp{r}} - \frac{\n{s} \times \n{r}}{\alp{s} \times \alp{r}}
\end{equation}
\begin{equation}\label{eq:5}
\begin{split}
\frac{Faulty}{\n{s} \times \n{r}}   & < \frac{1}{\alp{s}} + \frac{1}{\alp{r}} - \frac{1}{\alp{s} \times \alp{r}} \\
                                    & = \frac{\alp{r} + \alp{s} - 1}{\alp{s} \times \alp{r}}
\end{split}
\end{equation}
 $\frac{\alp{r} + \alp{s} - 1}{\alp{s} \times \alp{r}} <= \frac{3}{4}$ for $a_s,a_r>=2$. When substituted in Equation~\ref{eq:5} gives $\frac{Faulty}{\n{s} \times \n{r}} <= \frac{3}{4}$.
 $\frac{Faulty}{\n{s} \times \n{r}}$ is also the probability of selecting a faulty pair ($p_{fail}$).
So, the probability of selecting $q$ faulty pairs for resends is:
\begin{equation}
 (\frac{Faulty}{\n{s} \times \n{r}})^q  <= (\frac{3}{4})^q
\end{equation}
Solving for $q$ gives $q = \lceil \log_{\frac{3}{4}}{p_{fail}} \rceil$.
For $p_{fail} = 1 \times 10^{-9}$, we need at most $q = 72$ resends.
\end{proof}

\subsection{Pseudocode}
\label{subsec:pseudocode}

In Figure~\ref{fig:picsou-code}, we present the psuedocode for \Scrooge{}.
\textbf{Note:} This pseudocode does not include:
\begin{enumerate}
    \item Sending parallel acknowledgments (Section 4.2: \textit{Handling Failures} — The pitfalls of sequential recovery)
    \item Sending/receive scheduling based on stake (Section 5: \textit{Weighted RSMs Stake})
    \item The optional garbage collection (Section 4.3: \textit{Garbage Collection})
\end{enumerate}
Please refer to the preceding sections in this paper for a description of these techniques or consult the open-source C++ implementation at: \url{https://github.com/gupta-suyash/BFT-RSM}

\begin{figure*}[t]
\begin{lstlisting}[language=Python]
def transmit_message(msg, seq_num, cert):
    send_dest = message_scheduler.destination(seq_num, cur_node_id)

    # Ignore messages not handled by this node
    if send_dest == None:
        return
    # Send immediately or store for later
    if send_dest.resend_number == 0:
        payload = ((msg, seq_num, cert), local_ack, cur_node_id)
        send(first_send.destination, payload)
    else:
        pending_sends += required_sends

def receive_foreign_message(msg, seq_num, cert, sender_ack, sender_id):
    # Ignore invalid messages
    if is_cert_invalid((msg, seq_num), cert):
        return

    # Update local state
    ack.update(seq_num)
    quack.update(sender_ack)
    deliver(msg, seq_num)
    
    # Resend delayed messages, GC delivered messages
    for send in pending_sends:
        if quack.contains(send.seq_num):
            pending_sends.erase(send)
        elif quack.repeat_num(send.seq_num) >= send.resend_number:
            send(send.destination, send.payload)

    # Rebroadcast immediately or store for later
    rb_num = message_scheduler.rebroadcast_number(seq_num, cur_node_id)
    if rb_num == 0:
        local_broadcast(msg, seq_num, cert, ack, cur_node_id)
    elif rb_num > 0:
        pending_broadcasts.append((msg, seq_num, cert), rb_num)

def receive_local_message(msg, seq_num, cert, sender_ack, sender_id):
    # Ignore invalid messages
    if is_cert_invalid((msg, seq_num), cert):
        return

    # Update local state
    local_quack.update(sender_id, sender_ack)
    ack.update(seq_number)
    deliver(msg, seq_num)

    # Rebroadcast delayed local messages, GC delivered local messages
    for broadcast in pending_broadcasts:
        if local_quack.contains(broadcast.seq_num):
            pending_broadcasts.erase(broadcast)
        elif local_quack.repeat_num(broadcast.seq_num) >= broadcast.rebroadcast_num:
            local_broadcast(msg, seq_num, cert, ack, cur_node_id)
\end{lstlisting}
\caption{\Scrooge{} Psuedocode}
\label{fig:picsou-code}
\end{figure*}